%% file: RR-7812.tex
\documentclass[a4paper,twoside]{article}
\usepackage[latin1]{inputenc} 
\usepackage[T1]{fontenc} 
\usepackage{RR}
\usepackage{hyperref}

\usepackage{amsmath,amssymb}
\usepackage{arydshln}		
\usepackage{graphicx}

\usepackage{url}
\urldef{\mailsa}\path|{konstantinos.karanasos, asterios.katsifodimos,ioana.manolescu}@inria.fr|
\urldef{\mailsb}\path|spyros@zoupanos.net|

\usepackage{color}
\usepackage{soul}
\sethlcolor{red}

\newcommand{\anc}{{\prec}\mskip-8mu{\prec}\mskip2mu}
\newcommand{\precc}{{\prec}\mskip-8mu{\prec}\mskip2mu}
\newcommand{\kwd}[1]{\textsf{\small #1}}

\ifx \shrinkhdr \undefined
  \newcommand{\shrinkhdr}{\vspace{-1.6mm}}
\fi
\ifx \shrinkhdrafter \undefined
  \newcommand{\shrinkhdrafter}{\vspace{-1.02mm}}
\fi
\newcommand{\Subsubsection}[1]{\shrinkhdr\subsubsection{#1}\shrinkhdrafter}

\newcommand{\Subsection}[1]{\shrinkhdr\subsection{#1}\shrinkhdrafter}
\newcommand{\Section}[1]{\shrinkhdr\shrinkhdr\section{#1}\shrinkhdrafter}

\usepackage{tikz}
\tikzset{
	child/.style={thick,draw},
	descendant/.style={thick,double,draw,opacity=1},
	value join/.style={thick,dashed,draw,bend right=30,opacity=1},
	transparent edge/.style={draw,opacity=0},
	pname/.style={draw},
	plain/.style={opacity=1}
}

\RRNo{7812}
\RRdate{November 2011}
\RRauthor{Konstantinos Karanasos\thanks[sfn]{INRIA Saclay--\^Ile de France and LRI, Universit\'e Paris Sud-11}\and Asterios Katsifodimos\thanksref{sfn} \and\\
Ioana Manolescu\thanksref{sfn} \and Spyros Zoupanos\thanksref{sfn}\thanks{Max-Planck Institut f\"{u}r Informatik, Saarbr\"{u}cken, Germany}}

\authorhead{Karanasos, Katsifodimos, Manolescu \& Zoupanos}
\RRtitle{La plate-forme ViP2P: vues XML en pair-\`a-pair}
\RRetitle{The ViP2P Platform: XML Views in P2P}
\titlehead{The ViP2P Platform: XML Views in P2P}
\RRresume{Les grands volumes de donn\'ees XML disponibles sur le Web, produites par les organisations ou individus posent des d\'efis importants pour la gestion efficace de donn\'ees. Ce travail est situ\'e dans le contexte de la gestion de grands volumes de documents XML, dans un r\'eseau d\'ecentralis\'e, distribu\'e, pair-\`a pair, qui s'appuye sur une table de hashage distribu\'ee (ou DHT). Dans ce rapport, nous pr\'esentons ViP2P ({\em v}ues en {\em p}air-\`a-{\em p}air), une plateforme distribu\'ee pour le partage de documents XML s'appuyant sur un r\'eseau de type DHT. Au c\oe ur de ViP2P sont des vues mat\'erialis\'ees distribu\'ees. Celles-ci sont d\'efinies par n'importe quel pair, sous la forme de requ\^etes XML. D\`es que des donn\'ees XML publi\'ees par un pair quelconque  correspondent aux d\'efinitions des vues, ces donn\'ees seront utilis\'ees pour contribuer au contenu des vues. ViP2P fournit deux sc\'enarios d'\'evaluation de requ\^etes sur des documents XML. Il existe d'abord un mode ``souscription'', o\`u une requ\^ete enregistr\'eee dans le syst\`eme re\c{c}oit des r\'eponses de fa\c{c}on incr\'ementale, lorsque  des donn\'ees que l'on vient de publier contribuent aux r\'esultats. En deuxi\`eme lieu, une requ\^etes peut \^etre \'evalu\'ee uniquement \`a partir des donn\'ees d\'ej\`a publi\'ees, en r\'e\'ecrivant la requ\^ete \`a l'aide des vues mat\'erialis\'ees. Nous avons test\'e ViP2P d\'eploy\'e dans des r\'eseaux distribu\'es de plusieurs centaines d'ordinateurs. Dans ce rapport, nous pr\'esentons son architecture, ses principaux algorithmes, et d\'emontrons son efficacit\'e et son passage \`a l'\'echelle par une s\'erie d'exp\'eriences. Les r\'esultats de nos mesures d\'emontrent la robustesse de ViP2P jusqu'\`a des volumes de donn\'ees, d\'ebit de dissemination de donn\'ees, et tailles de r\'eseau, allant au del\`a (jusqu'\`a plusieurs ordres de grandeurs)  des mesures pr\'ec\'edemment publi\'ees sur des syst\`emes comparables.}

\RRabstract{ The growing volumes of XML data sources on the Web or produced by enterprises, organizations etc. raise many performance challenges for data management applications. In this work, we are concerned with the distributed, peer-to-peer management of large corpora of XML documents, based on distributed hash table (or DHT, in short) overlay networks. We present ViP2P (standing for {\em Vi}ews in {\em P}eer-to-{\em P}eer), a distributed platform for sharing XML documents based on a structured P2P network infrastructure (DHT). At the core of ViP2P stand {\em distributed materialized XML views}, defined by arbitrary XML queries, filled in with data published anywhere in the network, and exploited to efficiently answer queries issued by any network peer. ViP2P allows user queries to be evaluated over XML documents published by peers in two modes. First, a long-running subscription mode, when a query can be registered in the system and receive answers incrementally when and if published data matches the query. Second, queries can also be asked in an ad-hoc, snapshot mode, where results are required immediately and must be computed based on the results of other long-running, subscription queries. ViP2P innovates over other similar DHT-based XML sharing platforms by using a very expressive structured XML query language. This expressivity leads to a very flexible distribution of XML content in the ViP2P network, and to efficient snapshot query execution. ViP2P has been tested in real deployments of hundreds of computers. We present the platform architecture, its internal algorithms, and demonstrate its efficiency and scalability through a set of experiments. Our experimental results outgrow by orders of magnitude similar competitor systems in terms of data volumes, network size and data dissemination throughput.}

\RRmotcle{pair-\`a-pair, XML, THD, execution de requ\^etes distribu\'ee, evaluation de requ\^etes en terme de vues}
\RRkeyword{P2P, XML, DHT, distributed query processing, view-based query evaluation}
\RRprojet{Leo}
\RCSaclay 

\begin{document}
\makeRR   

\input{introduction} \newpage
\input{stateart} \newpage
\input{platform} \newpage
\input{patterns} \newpage
\input{internals} \newpage
\input{experimentation} \newpage
\input{conclusion} \newpage

\tableofcontents

\bibliographystyle{abbrv}
\bibliography{the}

\end{document}

%% file: introduction.tex
\Section{Introduction}
The volumes of data sources available in the form of XML documents has exploded since the W3C's 1998 standard, and so have the languages, tools and techniques for efficiently processing XML data. The interest of distribution in this context is twofold. First, a distributed storage and processing network can accommodate data volumes going far beyond the capacity of a single computer. Second, as organizations and individuals interact more and more, sharing and consuming one another's information flows, it is often the case that (XML) data sources  are produced independently by several distributed sources. The set of producers and consumers of data related to a specific topic, e.g., IT journals, blogs and online bulletins, is not only distributed, but also dynamic: sources may join or leave the system, the set of information consumers or their topics of interest may also change in time etc. Thus, we are interested in the large-scale management of distributed XML data in a {\em peer-to-peer} (P2P) setting. To provide users with {\em precise, detailed and complete} answers to their requests for information, we adopt a database-style approach where such requests are formulated by means of a structured query language, and the system must return complete results. That is, if somewhere in the distributed peer network, an answer to a given query exists, the system will find it and include it in the query result. To achieve this, our goal is to build a P2P XML data management platform based on a distributed hash table (or DHT, in short~\cite{DBLP:conf/iptps/DabekZDKS03}).

In this setting, users may formulate two kinds of information requests. First, they may want to {\em subscribe to interesting data anywhere in the network}, and published before or after the subscription is recorded in the system. Our goal is to persist the subscriptions and ensure that results are eventually returned as soon as possible following the publication of a matching data source. This is in the spirit, e.g., of RSS feeds, but extended to a distributed network where the source from which interesting data will come is not a priori known.
Second, users may formulate {\em ad-hoc (snapshot)} queries, by which they just seek to obtain as fast as possible the results which have already been published in the network.  

The challenges raised by a DHT-based XML data management platform  are:

\begin{itemize}
\vspace{-2mm}
\item building a {\em distributed resource catalog}, enabling client producers and consumers to ``meet'' in the virtual information sharing space; such a catalog is needed both for subscription and ad-hoc queries,
\item efficiently distributing the data of the network to the consumers that have subscribed to it and
\item providing {\em efficient distributed query evaluation algorithms} for answering ad-hoc queries fast.

\end{itemize}


In this paper, we present ViP2P, standing for {\em Views in Peer-to-Peer}, a distributed P2P platform for sharing Web data, and in particular XML data. ViP2P is built on top of a structured P2P network infrastructure, and it allows each peer in the network to share data with all the other peers. Data sharing in ViP2P is twofold. First, each network peer can ask long-running queries which are treated as subscriptions, that is, they receive results if and when a document published in the system matches such queries. Second, once results are stored for such a subscription, they are treated as {\em materialized views} based on which subsequent ad-hoc queries can be processed with snapshot semantics, i.e., based only on the data already published in the network. Given such an ad-hoc query, a ViP2P peer looks up the ViP2P network for relevant materialized views, runs an algorithm for equivalently rewriting the query, identifies and evaluates a distributed query evaluation plan which, based on the views, computes exactly the results of the query on the data published in the system prior to the query. ViP2P thus fills two kinds of needs: ($i$)~disseminating information in a timely fashion to subscriber peers and ($ii$)~re-using pre-computed results to process ad-hoc queries efficiently on the existing data only. 

A critical issue when deploying XML data management applications on a DHT is the division of tasks between the DHT and the upper layers. The DHT software running on each machine allows peers to remain logically connected to each other and to look up data based on search keys: a small set of simple, light-weight operations. In contrast, powerful XML data management requires complex languages (such as the W3C's XPath and XQuery standards), and scalable algorithms to cope with complex processing and large data transfers (known to raise performance issues in any distributed data management setting). 

Experience with our previous DHT-based XML data management platform  KadoP~\cite{DBLP:conf/icde/AbiteboulMPPS08} has taught us to load the DHT layer {\em as little as possible}, and keep the heavy-weight query processing operations in the data management layer and outside the DHT.  This has enabled us to build and efficiently deploy a system of important size (70.000  lines of Java code), which, as we show, scales on up to 250 computers in a WAN, and hundreds of GBs of XML data. ViP2P improves over the state of the art in DHT-based XML data management, since: ($i$)~it is one of the very few systems actually implemented (together with~\cite{DBLP:conf/icde/AbiteboulMPPS08,10.1109/TKDE.2009.26}, and opposed to prototypes built on DHT simulators), ($ii$)~is shown to scale on data volumes that are orders of magnitude beyond the cited competitor systems and  ($iii$)~has the most expressive XML query language, and the most advanced capabilities of re-using previously stored XML results, among all similar existing platforms~\cite{DBLP:conf/icde/AbiteboulMPPS08,DBLP:journals/dke/BonifatiC06,DBLP:conf/widm/BonifatiMCJ04,DBLP:conf/vldb/GalanisWJD03,PitouraSIGMODRecord2005,PitouraSIGMOD2008,10.1109/TKDE.2009.26}.

ViP2P  is part of a family of systems aiming at efficient management of XML data in structured peer-to-peer networks~\cite{DBLP:conf/icde/AbiteboulMPPS08,DBLP:journals/dke/BonifatiC06,DBLP:conf/widm/BonifatiMCJ04,DBLP:conf/vldb/GalanisWJD03,PitouraSIGMODRecord2005,PitouraSIGMOD2008,10.1109/TKDE.2009.26,DBLP:conf/icde/RaoM09}. The contributions of this work, with respect to the existing systems, are as follows:

\begin{itemize}
\item We present a {\em complete architecture for query evaluation, both in continuous (subscription) and in snapshot mode}. This architecture enables the efficient dissemination of answers to tree pattern queries (expressed in an XQuery dialect) to peers which are interested in them, regardless of the relative order in time between the data and the subscription publication. As in~\cite{PitouraSIGMOD2008}, it also allows to efficiently answer queries in snapshot mode, based on the content of the existing views materialized in the network, but using more expressive views, queries and rewritings. 
\item We have fully implemented our architecture (about 250 classes and 70.000 lines of Java code), on top of the FreePastry~\cite{FreePastry} P2P infrastructure. 
We present a {\em comprehensive set of experiments performed in a WAN}, showing that ($i$)~the performance of a fully deployed large-scale distributed system (and in particular a DHT-based XML management platform) is determined by many parameters, beyond the network size and latency which can be set in typical P2P network simulators and ($ii$)~the ViP2P architecture scales to several hundreds of peers and hundreds of GBs of XML data, both unattained in previous works.
\end{itemize}

The paper is organized as follows. Section~\ref{sec:state} surveys the state of the art in managing XML data in DHT networks. Section~\ref{sec:platformOverview} introduces the ViP2P architecture via an example and describes its main modules.  Section~\ref{sec:rew-outline} presents the query and view language, as well as query rewriting in ViP2P, while Section~\ref{sec:viewIndexingAndRetrieval} concentrates on the materialization, indexing and look-up of materialized views, at the core of the platform. In Section~\ref{sec:exp}, we present a set of experiments analyzing the performance of ViP2P data management in a variety of settings and demonstrating its scalability, then we conclude.

%% file: stateart.tex
\Section{State of the art}
In this section we present the current state of the art in XML data management over P2P networks. In Section~\ref{sec:struct-vs-unstruct} we focus on the differences of structured and unstructured P2P networks and the reasons behind our choice to use a structured P2P network for building our platform. In Section~\ref{sec:xml-on-dhts} we present our closest competitor works focusing on the management of XML data over structured DHT networks. Section~\ref{sec:platforms-vs-simulations} stresses the challenges of distributed XML data management in a real, deployed platform as opposed to simulations. Finally, in Section~\ref{sec:previous-pubs}, we present earlier publications of the ViP2P platform.

\label{sec:state}
\Subsection{Structured vs. unstructured P2P networks} 
\label{sec:struct-vs-unstruct}

Peer-to-peer content sharing platforms can be broadly classified in two groups. Unstructured peer-to-peer networks allow arbitrary connections among peers, that is, each peer may be connected to (or aware of the existence of)  one or more network peers of its choice. Such network structure typically mimics some conceptual proximity between peers interested, for instance, in similar topics. Structured peer networks, on the other hand, impose the set of connections among peers. A survey of (structured and unstructured) P2P XML sharing platforms reflects the state of the art and open issues as of 2005~\cite{PitouraSIGMODRecord2005} and a more recent survey of XML document indexing and retrieval in P2P networks can be found in~\cite{AbererP2PSurvey}.

The different network structures impact the way in which searches (or queries) can be answered in the network. Thus, in unstructured networks, queries are forwarded from each peer to its set of known peers (or neighbors) and answers are computed gradually as the query reaches more and more peers. For instance, in~\cite{XPeer} peers are logically organized into clusters that are formed on a document schema-similarity basis. The superpeers of the network are organized to form a tree, where each superpeer hosts schema information about its children. When a query arrives it is forwarded to the superpeers. Every superpeer performs location assignment: it examines the schemas of the documents of its children to detect which peers could possibly contribute results to the query. After the contributing peers have been located, the peer that originally posed the query builds a location aware algebraic plan and ships the corresponding subqueries to their respective peers. The results are then retrieved from each peer and the original query is evaluated by performing operations such as joins over the subquery results.

It is easy to see that if query answers reside on a peer very far (in terms of peer connections) from the peer where the query originated, this may lead to numerous messages and a long query response time. To improve the precision, performance and recall of query answering in this context, many approaches have been proposed, from the earliest~\cite{DBLP:conf/icde/YangG03} to the very recent~\cite{DBLP:conf/p2p/DedzoeLAV10}, to name just a few.

In contrast, structured networks (and their best-known representatives, distributed hash tables or DHTs, in short~\cite{DBLP:conf/iptps/DabekZDKS03}) provide a simple distributed index functionality implemented jointly by all the peers. The simplest DHT interface provides {\em put(key, value)} and {\em get(key)} operations allowing the storage of (key, value) pairs distributed over all the network peers. More advanced DHT structures also allow range searches of the form {\em get(key range)}, such as Baton~\cite{DBLP:conf/sigmod/JagadishOTVZ06,DBLP:conf/vldb/JagadishOV05} or P2PRing~\cite{DBLP:conf/sigmod/CrainiceanuLMGS04,DBLP:conf/sigmod/CrainiceanuLMGS07}. In a DHT, to answer a {\em get} request, a bounded number of messages are exchanged in the network,  typically in $O(log_2(N))$, where $N$ is the number of network peers.

In this work, we consider the setting of a structured network, based on a DHT, and design an efficient platform for {\em XML query processing in large scale networks, based on P2P XML materialized views}. 
The main difference between most of the existing platforms and ViP2P is that our system addresses the whole processing chain involved in evaluating queries, as opposed to only locating the interesting documents and shipping the query to those peers for evaluation. The latter approach may, in some cases, require numerous messages at query evaluation time and possibly increased response times. ViP2P, in contrast, considers the complete chain of query processing based on materialized views incrementally built in the network. This enables answering queries by contacting only a few peers and possibly re-using complex pre-computed results, stored in the views.

\Subsection{XML data management based on DHTs}
\label{sec:xml-on-dhts}
The first DHT-based platform for XML content sharing was described in~\cite{DBLP:conf/vldb/GalanisWJD03}. This  work proposed a framework for indexing XML documents, based on the parent-child element paths appearing in the document. Processing a query involves ($i$)~extracting from the query a set of paths which could serve as lookup keys, ($ii$)~obtaining via {\em get} calls the IDs of all documents matching the paths, ($iii$)~shipping the query to all the peers holding such documents and ($iv$)~retrieving the results at the query peer. The approach carries some imprecision in the case of queries featuring the descendant axis (\kwd{//}) or tree branches. For instance, the query \kwd{/a[b]/c} could be forwarded to documents in which the paths \kwd{/a/b} and \kwd{/a/c} occur, but the tree pattern \kwd{/a[b]/c} does not occur. A very similar approach to DHT-based XML indexing by parent-child paths is taken in~\cite{DBLP:conf/otm/SkobeltsynHA05}.

The above discussion illustrates a common aspect in DHT-based content management platforms: imprecision in the indexing method leads to more peers being contacted to process a given query. A previous work on managing relational data based on DHTs~\cite{DBLP:conf/vldb/LooHHSS04} has shown that intensive messaging at query time may seriously limit scaling. Therefore, index precision is generally a desirable feature. 

The work described in~\cite{DBLP:journals/dke/BonifatiC06,DBLP:conf/widm/BonifatiMCJ04} considers the setting where XML documents are divided in fragments distributed among several peers. Each fragment is assigned as identifier the parent-child label path going from the document root to the root of the fragment, and subsequently, fragments are indexed in the DHT by their identifiers. The system uses a particular DHT which can handle {\em prefix queries}, and thus allows locating XML fragments for which a prefix of the path from the root to the fragment is known. Processing linear queries using only the child axis is simple, however, simple queries using the descendant axis, such as the query \kwd{//a}, need to be forwarded to all the network peers. 

The KadoP system~\cite{DBLP:conf/icde/AbiteboulMPPS08} indexes XML documents at fine granularity. Thus, for any element name \kwd{a}, a network peer is in charge of storing the identifiers (or IDs, in short) of all \kwd{a}-labeled elements from all the documents in the network. The IDs reflect the position of the elements in the respective documents. Therefore, any tree pattern query can be answered by retrieving the list of IDs corresponding to each tree pattern node, and combining these lists via a holistic twig join~\cite{DBLP:conf/SIGMOD/BrunoKS02}. This indexing model has very high precision, since the output of the holistic twig join includes exactly the documents matching the query. However, the index is much more voluminous than in previous proposals~\cite{DBLP:journals/dke/BonifatiC06,DBLP:conf/widm/BonifatiMCJ04,DBLP:conf/vldb/GalanisWJD03,DBLP:conf/otm/SkobeltsynHA05}, highlighting the severe limitations {\em in terms of volume of the (key, value) pairs} of the DHT index. Several optimizations in the index structure were introduced in~\cite{DBLP:conf/icde/AbiteboulMPPS08}, based on which the KadoP platform was tested on hundreds of peers and 1GB of data.

More recently,  the {\em psiX} system~\cite{DBLP:conf/icde/RaoM09,10.1109/TKDE.2009.26} proposed an XML indexing scheme based on document summaries, corresponding to the  backward simulation image of the XML documents (if a DTD is available, summaries can also be built based on the DTD). An algebraic signature is associated to each summary and to each query. When a query arrives, the algebraic query signature is used to look up in a holistic fashion all document signatures matching the query. The precision of this indexing scheme improves over KadoP~\cite{DBLP:conf/icde/AbiteboulMPPS08} by a better treatment of wildcard ($*$) nodes, which KadoP ignores for the most part of query processing. From the matching summaries, one can identify the concrete corresponding documents, and then push query evaluation to the peers hosting the documents. The approach is implemented over the Chord DHT and shown to be effective by experiments on up to 11 peers in the PlanetLab network. 

The main difference between the works described in~\cite{DBLP:conf/vldb/GalanisWJD03,DBLP:conf/icde/RaoM09,10.1109/TKDE.2009.26,DBLP:conf/otm/SkobeltsynHA05} and our work lies in the approach taken for query processing. These works, of which {\em psiX}~\cite{10.1109/TKDE.2009.26} can be considered the most advanced, are only concerned with locating the documents relevant for a query. In contrast, \cite{DBLP:journals/dke/BonifatiC06,DBLP:conf/widm/BonifatiMCJ04}, KadoP~\cite{DBLP:conf/icde/AbiteboulMPPS08} and the ViP2P platform presented here address the P2P XML query processing problem as a whole. They re-distribute data in the P2P network in order to prepare for the evaluation of future queries. KadoP distributes a tag index over the peers independently of the data and the queries, which can be seen as a ``one size fits all'' approach. ViP2P allows individual peers to choose the particular queries of interest for them, expressed in a rich tree pattern dialect (or, equivalently, a useful XQuery subset) and then allows exploiting the stored results of such queries as views for rewriting future queries. An ongoing development of ViP2P~\cite{LiquidXML} focuses on automatically choosing the views to materialize on each peer in order to improve observed query processing performance. Thus, going beyond the problem of {\em locating} relevant documents, ViP2P aims at making the most out of the existing network storage and processing capacity in order to {\em evaluate queries} most efficiently to the peers that need them.

Closer in spirit to our work is the cooperative XPath caching approach described in~\cite{PitouraSIGMOD2008}, where peers can store results of a (peer-chosen or system-imposed) XPath query. The definitions of these stored queries (or views) are indexed in the network, enabling subsequent queries to be rewritten and answered based on these views. ViP2P is more general, since ($i$)~our view and query language is an XQuery dialect with many returning nodes, as opposed to the simple XPath subset in~\cite{PitouraSIGMOD2008} and ($ii$)~our approach allows to rewrite a query based on {\em several} views, whereas~\cite{PitouraSIGMOD2008} can only exploit one view for one query. 

DHT-based XML indexing methods~\cite{DBLP:conf/icde/AbiteboulMPPS08,DBLP:journals/dke/BonifatiC06,DBLP:conf/widm/BonifatiMCJ04,DBLP:conf/vldb/GalanisWJD03,DBLP:conf/icde/RaoM09,10.1109/TKDE.2009.26,DBLP:conf/otm/SkobeltsynHA05} are {\em complete}, i.e., for each query, based on the index, all relevant answers can be computed and returned.  In ViP2P and~\cite{PitouraSIGMOD2008}, peer-chosen views replace the compulsory index fragments assigned by the network to each peer. Thus, it is possible that some queries cannot be processed due to the lack of appropriate views. Our focus in ViP2P is on efficiently building and exploiting pre-computed query results under the form of materialized views. To guarantee completeness, our approach can be coupled with an efficient and compact document-level index, such as {\em psiX}~\cite{10.1109/TKDE.2009.26}, on which to fall back when no suitable views are found for a given query.

We conclude our  analysis by considering the {\em granularity} or level of detail used to index XML, i.e., the granularity of the keys inserted in the DHT. Element labels (or label paths, or document summaries) have been often used. However, this does not allow efficiently locating documents which satisfy specific {\em value or keyword} search conditions, such as e.g., \kwd{//item[price=\$45]} or \kwd{//item[contains(.,'cam} \kwd{era')]}. Indexing by keywords or text nodes increases index precision but also significantly increases the index size, since there are many more keywords in an XML document than distinct tags.  Therefore, the approaches of~\cite{DBLP:journals/dke/BonifatiC06,DBLP:conf/widm/BonifatiMCJ04,DBLP:conf/vldb/GalanisWJD03,DBLP:conf/icde/RaoM09,10.1109/TKDE.2009.26,DBLP:conf/otm/SkobeltsynHA05} cannot be easily extended to support keyword search and preserve their scalability. A {\em value summary framework} is proposed in~\cite{DBLP:conf/vldb/GalanisWJD03} to index element values by trading off precision for index space. KadoP~\cite{DBLP:conf/icde/AbiteboulMPPS08} indexes all keywords just like element labels, and proposes  index-level optimization techniques to cope with important scale-related problems. ViP2P allows keyword and value conditions both in the materialized views and in the queries.

\Subsection{Managing XML on a DHT: platforms vs. simulations} 
\label{sec:platforms-vs-simulations}
Developing distributed systems, and in particular a P2P platform, requires significant efforts. This may be a reason why many previous works in this area validate their techniques based on {\em simulated peer networks}, where a single computer runs an analytical model configured to simulate a given network size. Our INRIA team has invested significant manpower (of the order of 70 man $\times$ month  by now) developing the KadoP and then the ViP2P platforms. Our effort has taught us that many architecture and engineering problems arise due to the mismatch between the initial DHT goals (maintaining large dynamic networks connected and providing minimal messaging), and the data-intensive operations required by indexing, storing, and querying large volumes of XML data. We have addressed these problems in ViP2P by careful architecture and engineering, and report in this paper {\em experiments at a scale (in peers deployed over a WAN, and in data size) unattained so far by any other platform}. Thus, KadoP~\cite{DBLP:conf/icde/AbiteboulMPPS08} scales up to 1 GB of data over 50 computers peers, psiX~\cite{10.1109/TKDE.2009.26} used 262 MBs of data and 11 computers, and in this paper we report on sharing up 160 GB of data over up to 250 computers (in all cases, the computers were distributed in a WAN). 

\Subsection{Previous publications on ViP2P} 
\label{sec:previous-pubs}
A first version of the platform was described in an informal setting (no proceedings) in an international workshop~\cite{DataX} and a national conference~\cite{BDA2009}. These works used a more restricted query language than we consider here, and described early experiments on a platform which has been much improved since. 
Two ViP2P applications have lead to demonstrations: P2P management of RDF annotations on XML documents~\cite{DBLP:conf/icde/KaranasosZ10} and adaptive content redistribution~\cite{LiquidXML}. The details of view-based query rewriting in ViP2P are described in a separate paper~\cite{rewritingICDE2011}. They can be seen as orthogonal to the architecture and performance issues described here.

%% file: platform.tex
\Section{ViP2P platform overview}
\label{sec:platformOverview}

XML data flows in ViP2P can be summarized as follows. XML documents are published independently and autonomously by any peer. Peers can also formulate subscriptions, or long-running queries, potentially matching documents published before, or after the subscriptions. The results of each subscription query are stored at the respective peer, and the definition of the query is indexed in the peer network. Finally, peers can ask ad-hoc queries, which are answered in a snapshot fashion (based on the data available in the network so far) by exploiting the existing subscriptions, which can be seen as materialized views. We detail the overall process via an example in Section~\ref{sec:sys-overview}. We then proceed to describe the ViP2P modules implementing it in Section~\ref{sec:archi}.

\begin{figure}[t!]
\begin{center}
\includegraphics[width=0.75\columnwidth]{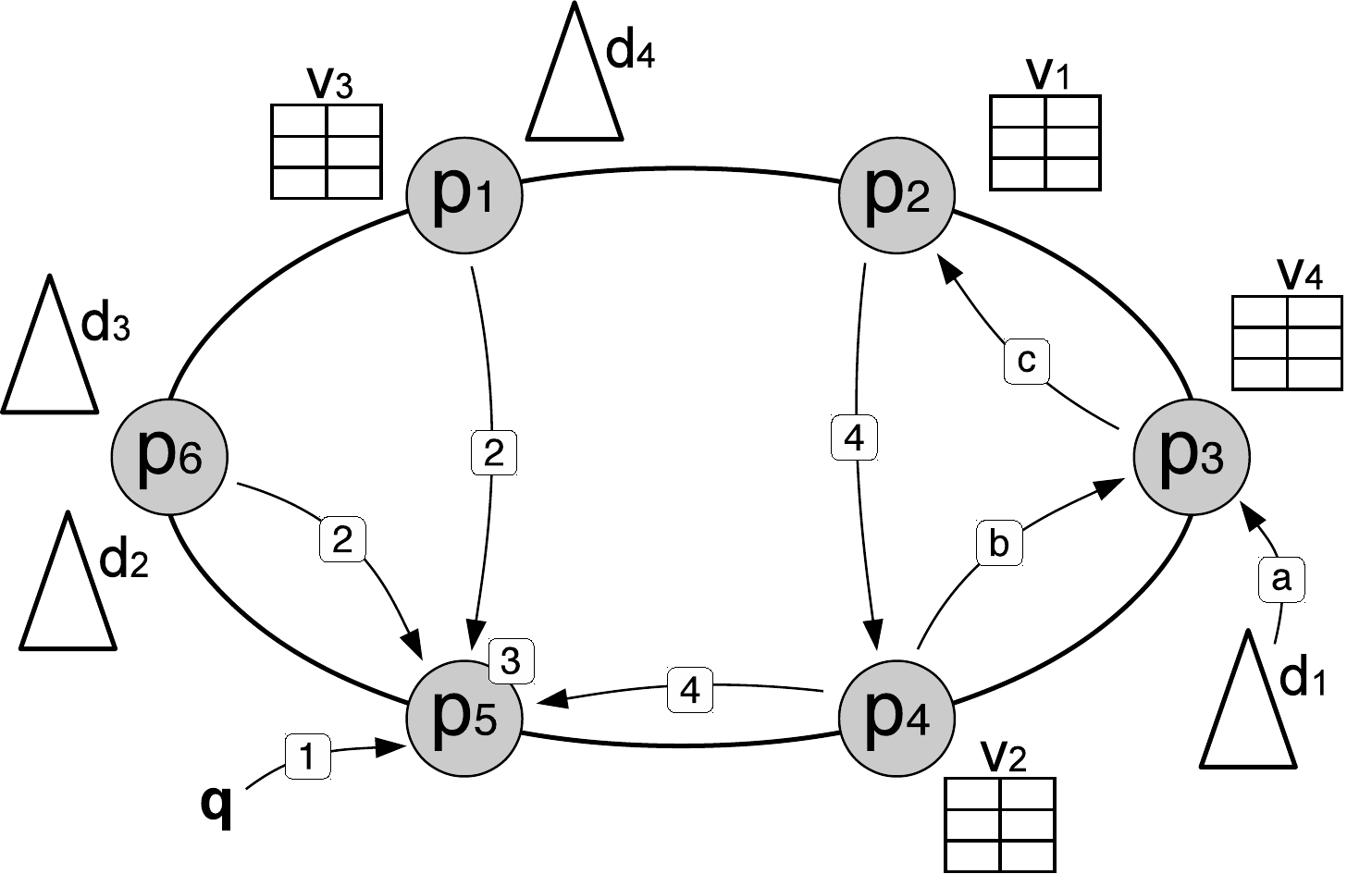}
\caption{System overview.\label{fig:ring}}
\end{center}
\end{figure}

\input{basic-platform}

\Subsection{ViP2P peer architecture}
\label{sec:archi}

\begin{figure}[t!]
	\begin{center}
		\includegraphics[width=0.85\columnwidth]{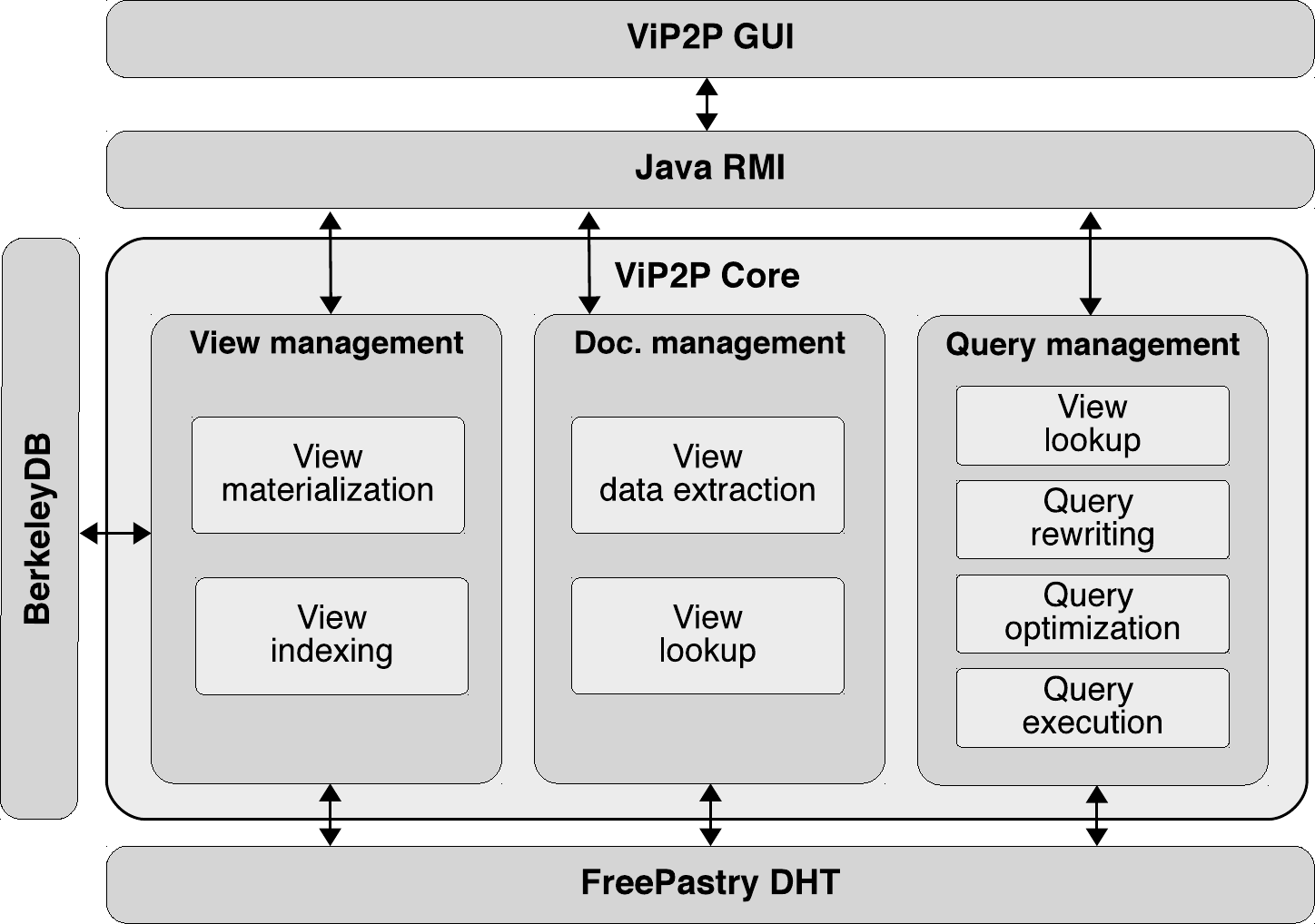}
        \end{center}
\vspace{-4mm}
\caption{Basic architecture of a ViP2P peer.\label{fig:vip2pPeer}}
\end{figure}

We now present the main modules of ViP2P peers as well as their functionalities and  interaction, outlined in Figure~\ref{fig:vip2pPeer}. The {\em ViP2P Core} box includes the main modules, whereas boxes located outside  \emph{ViP2P Core} are independent external subsystems that interact with ViP2P.

\Subsubsection{External Subsystems}
\label{sec:ext-subsystems}
\vspace{1mm}
\noindent{\em \bf FreePastry DHT}~\cite{FreePastry} provides the underlying DHT layer on which ViP2P is built. FreePastry is an open-source implementation of Pastry~\cite{Pastry}, an efficient, self-organizing and fault-tolerant overlay network. Pastry provides efficient request routing, deterministic object location, and load balancing. ViP2P nodes index and lookup view definitions on FreePastry's DHT during the view materialization and query processing.

\vspace{1mm}
\noindent{\em \bf Java RMI} is used for all large data transfers. Previous work~\cite{DBLP:conf/icde/AbiteboulMPPS08} has shown that the DHT communication primitives were not suitable for such transfers, since ($i$)~the DHT {\em get} and {\em put} operations are blocking, that is, data sent via the DHT becomes available at the receiver only when it has been completely received and ($ii$)~message queues in the DHTs overflow easily even after tuning, in which case the DHT peers re-send them, which further clogs the DHT communication pipes. Beyond the degradation of performance, such message overflows are annoying because a peer that is too busy trying to re-send data, may skip sending the regular ``ping'' to his neighbors to signal that it is still alive. Then, the neighbors suspect the peer is down, this triggers further loss of messages etc.

For all these reasons, we have decided to split inter-peer communication in two categories. The DHT is used to efficiently send small messages, typically to index and look up view definitions.  We use RMI (which we were able to fine-tune by writing efficient custom serialization/de-serialization methods, properly controlling concurrency at the send and receiver side etc.)  to send larger messages containing view tuples, when views are materialized and queried. We also applied specific techniques to reduce the space occupancy of transmitting tuples. Thus, a document ID (or URI) often appears many times in a view, as many times as there are view tuples obtained from that document. Since the URIs are quite large, they make up an important part of the document data. We use dictionary-based encoding of the document URIs, i.e., the tuple sender dynamically builds a dictionary of all document URIs and sends partial dictionaries with each tuple packet, to enable decoding on the receiver side. One could perhaps improve performance even further by coding data-intensive communications at a lower level (e.g. using plain sockets), but the improvements attained by our way of utilizing RMI are already very significant.

\vspace{1mm}
\noindent{\em \bf BerkeleyDB} Within each peer, view tuples are efficiently stored into a native store that we built using the Berkeley DB~\cite{BDB}  library. It provides the routines to store, retrieve and sort entries, while guaranteeing ACID transactions when view data are written and read concurrently. 

\vspace{1mm}
\noindent{\em \bf The GUI} facilitates the control and inspection of each peer, enabling users to 
publish views and/or pose queries. Screenshots of the ViP2P GUI, along with other information, can be found on the ViP2P website\footnote{http://vip2p.saclay.inria.fr/}.

\vspace{1mm}
\noindent We now move to describing the core modules.

\Subsubsection{Document management module} This module is responsible for looking up for views to which the peer's documents may contribute, extracting the data from the documents and sending it to the respective consumers. 

\vspace{1mm}
\noindent{\textbf{View definition lookup}} When a new document is published by a peer, the \emph{view lookup} module at this peer first, looks up in the DHT the definitions of the views to which the document may contribute data, and then passes these views definitions to the \emph{view data extraction module}.

\vspace{1mm}
\noindent{\textbf{View data extraction}} Given a list of view definitions, the \emph{view data extraction module} at a publisher peer extracts from the document the tuples matching each view, and ships them, in a parallel fashion, to the different consumers. The view data extractor is capable of simultaneously matching several views on a given document. Thus, the corresponding tuples are extracted during a single traversal of the document. The extractor maintains a thread pool for setting up RMI communications for shipping tuples to the consumers. As our experiments show in Section~\ref{sec-exp-materialization}, this parallel tuple sending significantly reduces the time needed to materialize the views.

\begin{figure}[t!]
	\begin{center}
		\scalebox{1.5}{
		\begin{tikzpicture}
			\draw (0cm, -0.8cm) -- (0cm, 1.4cm);
			\draw (-3cm, -0.8cm) -- (-3cm, 1.4cm);
			\draw (0cm, 1.4cm) node[above=2pt] {\scalebox{0.8}{view~holder}};
			\draw (-3cm, 1.4cm) node[above=0.5pt] {\scalebox{0.8}{tuple~extractor}};
			\draw (-3.8cm, 1.2cm) node {\tiny tuples~ready};
			\draw (-1.5cm, 0.93cm) node[above=1pt,rotate=-5] {\tiny tuple-send~request};
			\draw [->] (-3cm, 1.2cm) -- (0cm, 0.9cm);
			\draw (1cm, 0.85cm) node {\tiny enqueue~request};
			\draw (-1.5cm, 0.5cm) node[above=1pt,rotate=5] {\tiny busy};
			\draw [<-] (-3cm, 0.4cm) -- (0cm, 0.8cm);
			\draw (-3.4cm, 0.4cm) node {\tiny sleep};
			\draw (1cm, 0.3cm) node {\tiny dequeue~request};
			\draw (-1.5cm, 0cm) node[above=1pt,rotate=5] {\tiny ready};
			\draw [<-] (-3cm, -0.1cm) -- (0cm, 0.3cm);
			\draw (-3.58cm, -0.1cm) node {\tiny wake~up};
			\draw (-1.5cm, -0.5cm) node[above=1pt,rotate=-5] {\tiny send~tuples};
			\draw [->] (-3cm, -0.2cm) -- (0cm, -0.6cm);
			\draw (0.8cm, -0.6cm) node {\tiny store~tuples};
		\end{tikzpicture}
		}
		\vspace{-3mm}
		\caption{Tuple-send/receive protocol use case between a tuple--sender and a view holder.\label{fig:tuplesendprotocol}}
		
        \end{center}
\end{figure}

\Subsubsection{View management module} This module handles view indexing and materialization. 

\vspace{1mm}
\noindent{\textbf{View indexing}} This module makes visible to all network peers the definitions of all the views declared in the ViP2P network (of course without broadcasting them, since most peers are typically not interested in all views).  When a new view is defined, the indexer inserts in the DHT (key,value) pairs used to describe it, based on one of the \emph{indexing strategies} that we will describe in Section~\ref{sec:digestAndRepository}. 

\vspace{1mm}
\noindent{\textbf{View materialization}} The \emph{view materialization} module receives tuples from remote publishers and stores them in the respective BerkeleyDB database. In a large scale, real-world scenario, thousands of documents might be contributing data to a single view. To avoid overload on its incoming data transfers, this module implements a back-pressure  \emph{tuple-send/receive protocol} which informs the publisher when the incoming tuple buffer is full at the consumer side. Thus, a publisher may have to wait until the consumer is ready to accept the tuples. This makes the most out of the available publisher-to-consumer bandwidth, all the while avoiding costly re-transmissions due to messages lost from overflowing queues.

Figure~\ref{fig:tuplesendprotocol} traces the tuple-send/receive protocol between a tuple extractor and a view holder. First the tuple extractor extracts the tuples and keeps them in memory being ready to ship them to the view holder. After that, it sends a tuple-send request to the view holder. In this example, the view holder is busy storing tuples (possibly sent by other tuple extractors), thus it enqueues the request and responds to the tuple extractor with a ``busy'' response. When the view holder is ready to accept the new set of tuples, it dequeues the request and informs the tuple extractor (via a ``ready'' message). Then, the tuple extractor ships the new tuples to the view holder, who finally stores them in the Berkeley database of the respective view. The view holder can serve multiple tuple-send requests concurrently. Our experiments (Section~\ref{sec-exp-materialization}) show how the concurrency can affect the time needed for a set of views to be materialized.

\Subsubsection{Query management module} 
A sequence of steps are required to evaluate queries, each performed by a dedicated module, as follows. 

\vspace{1mm}
\noindent{\textbf{View lookup}} This module, given a query, performs a lookup in the DHT network retrieving the view definitions that can be used to rewrite the query. 

\vspace{1mm}
\noindent{\textbf{Query rewriting}} This module takes a given ad-hoc query and a set of available view definitions and produces a logical rewriting plan which, evaluated on some views, produces exactly the results required by the query (algorithm detailed in~\cite{rewritingICDE2011} and illustrated in Section~\ref{sec:rew-outline}). 

\vspace{1mm}
\noindent{\textbf{Query optimization}} This module receives as input a logical rewriting plan which is output by the query rewriting module and translates it to an optimized physical plan. The optimization takes place both at the logical (join reordering, push selections and projections etc.) and physical (dictating the exact flow of data during query execution, selection of the appropriate physical operators etc) level.

\vspace{1mm}
\noindent{\textbf{Query execution}} This module provides a set of physical operators which can be executed by any ViP2P peer, implementing the standard iterator-based execution model~\cite{DBLP:conf/sigmod/Graefe90}. Since ViP2P is a distributed application, operators can be deployed to peers and executed in a remote manner. The query optimization module is the one to decide the parts of a physical plan that every peer executes.

Data exchange operators are an essential part of a distributed execution plan. To that end, ViP2P implements two data exchange operators: the {\em Send} and {\em Receive} operators that permit data exchange across peers. They are always used in pairs: whenever a data sender peer executes a {\em Send} operator, the data receiver executes its respective {\em Receive} operator. {\em Send} and {\em Receive} are implemented using asynchronous communication buffers (tuples are not sent through the network one by one but in buckets of specified size) and data is transferred via RMI. To reduce the transferred data volumes, document URIs (present in each view tuple to identify the document the tuple was extracted from) are compressed using a dictionary by the {\em Send} and decompressed by the {\em Receive} as described in Section~\ref{sec:ext-subsystems}. 

ViP2P implements the typical {\em Selection}, {\em Projection}, {\em Hash Join}, {\em Nested Loop Join} and {\em Merge Join} operators. Moreover, it uses the XML specific operators {\em Holistic Twig Join}~\cite{DBLP:conf/SIGMOD/BrunoKS02}, {\em Structural Ancestor Join} and {\em Structural Descendant Join}~\cite{DBLP:conf/icde/Al-KhalifaJPWKS02} performing structural joins based on the structural identifiers (IDs) of the incoming tuples. The {\em Navigation} operator corresponds to the logical navigation operator, described in Section~\ref{sec:rw}. Two sorting operators are available: an in-memory sort operator {\em Memory Sort}, and an external memory sort based on BerkeleyDB.


%% file: basic-platform.tex
\Subsection{ViP2P by example}
\label{sec:sys-overview}

A sample ViP2P instance over six peers is depicted in Figure~\ref{fig:ring} and we use it to base our presentation of the operations which can be carried in each peer. In the Figure, XML documents are denoted by triangles, whereas views are denoted by tables, hinting to the fact that they contain sets of tuples. More details on views and view semantics are provided in Section~\ref{sec:viewIndexingAndRetrieval}, but they are not required to follow the discussion here. For ease of explanation, we make the following naming conventions for the remainder of this paper:
\begin{itemize}
\item \textbf{Publisher} is a peer which publishes an XML document
\item \textbf{Consumer} is a peer which defines a subscription and stores its results (or, equivalently, the respective materialized view)
\item \textbf{Query peer} is a peer which poses an \emph{ad-hoc} query (to be evaluated over the complete ViP2P network).
\end{itemize}

\noindent Clearly, a peer can play any subset of these roles simultaneously or successively.

\Subsubsection{View publication} A ViP2P view is a \emph{long-running subscription query} that any peer can freely define.  The {\em definition} (i.e., the actual query) of each newly created view is indexed in the DHT network. For instance, assume peer $p_2$ in Figure~\ref{fig:ring} publishes the view $v_1$, defined by the XPath query $//bibliography//book[contains(.,'Databases')]$. The view requires all the books items from a bibliography containing the word `Databases'. ViP2P indexes $v_1$ by inserting in the DHT the following three (key, value) pairs: $(bibliography, v_1@p_2)$, $(book, v_1@p_2)$ and $('Databases', v_1@p_2)$. Here, $v_1@p_2$ encapsulates the structured query defining $v_1$, and a pointer to the concrete database at peer $p_2$ where $v_1$ data is stored. As will be shown below, all existing and future documents that can affect $v_1$, \emph{push} the corresponding data to its database.

Peers look up views in the DHT in two situations: when publishing documents, and when issuing ad-hoc queries. We detail this below.

\Subsubsection{Document publication}
When publishing a document, each peer is in charge of identifying the views within the whole network to which its document may contribute. For instance, in Figure~\ref{fig:ring} (step a), peer $p_3$ publishes the document $d_1$ (depicted in Figure~\ref{fig:xmlDoc}). Document $d_1$ contains data matching the view $v_1$ as it contains the element names $bibliography$ and $book$, as well as the word $'Databases'$. Peer $p_3$ extracts from $d_1$ all distinct element names and all keywords. For each such element name or keyword $k$, $p_3$ looks up in the DHT for view definitions associated to $k$, and, thus, learns about $v_1$ (step~b). In the publication example above, $p_3$ extracts from $d_1$ the results matching $v_1$; from now on, we will use the notation $v_1(d_1)$ to designate such results. Peer $p_3$ sends $v_1(d_1)$ to $p_2$ (step~c), which adds them to the database storing $v_1$ data.

\begin{figure}[t!]
\begin{center}
\scalebox{0.8}{\begin{tikzpicture}
[level distance=10mm]
\tikzstyle{level 1}=[sibling distance=3.2cm]
\tikzstyle{level 2}=[sibling distance=2cm]
\tikzstyle{level 3}=[sibling distance=1.4cm]
\node {$bibliography$}
	child {node {$book$}
		child {node {$title$}
			child {node[text width=2cm] {$Found. of$\\$Databases$}}
		} 
		child {node {$author$}
			child {node {$name$}
				child {node {$first$}
					child {node {$Serge$}}
				}
				child {node {$last$}
					child {node {$Abiteboul$}}
				}
			}
		}
		child {node {$year$}
			child {node {$1995$}}
		}
	}
	child {node {$\dots$}}
	child {node {$paper$}
		child {node {$title$}
			child {node[text width=1cm] {$AXML$\\$project$}}
		} 
		child {node {$author$}
			child {node {$name$}
				child {node {$first$}
					child {node {$Serge$}}
				}
				child {node {$last$}
					child {node {$Abiteboul$}}
				}
			}
		}
		child {node {$year$}
			child {node {$2008$}}
		}
	}
; 
\end{tikzpicture}}
\end{center}
\vspace{-5mm}
\caption{Sample XML document $d_1$.\label{fig:xmlDoc}}
\vspace{-2mm}
\end{figure}

A separate mechanism is needed for a view, say $v_x$, published after $d_1$ but having results in $d_1$. One possibility would be for the peer publishing $v_x$ to look up, among all the network documents, for those that could contain terms from $v_x$ and require them to contribute $v_x$ results.  The drawback is that this requires indexing  all documents on all terms, which may be wasteful since  a large part of published content may not be looked up frequently, or not at all.

Instead, ViP2P associates to each view an {\em interval timestamp}, corresponding to a time interval during which the view was published. Each peer having published a document $d$ must check the DHT for views published after $d$. To that effect, each peer performs regular lookups using as key, the time interval which has just passed. Thus, it retrieves the definitions of all the views published during that interval and contributes its data if it hasn't done it already.
\vspace{1mm}

\Subsubsection{Ad-hoc query answering}
ViP2P peers may pose {\em ad-hoc queries}, which must be evaluated immediately (from the previously published data). To evaluate such queries, a ViP2P peer looks up in the network for views which may be used to answer it. For instance, assume the query $q=//bibliography//book[contains(.,'Databases')]//author$ is issued at peer $p_5$ (step~1, in Figure~\ref{fig:ring}). To process $q$, $p_5$ looks up the keys $bibliography$, $book$, $'Databases'$ and $author$ in the DHT, and retrieves a set of view definitions (step~2), including that of $v_1$. Observe that $q$ can be rewritten as $v_1//author$; therefore, $p_5$ can answer $q$ just by retrieving and extracting $q$'s results out of $v_1$. A distinguishing feature of ViP2P (step~3) is its ability to combine {\em several} materialized views in order to rewrite a query (as we describe in Section~\ref{sec:rew-outline}). A query rewriting (a logical plan based on some views) is translated by the ViP2P query optimizer into a distributed physical plan, specifying which operators will be used and in which peers they will be executed. The ViP2P optimizer is responsible of selecting the most efficient physical plan, as this choice has a significant impact in the query execution time, especially in a distributed setting such as ours where network communication plays an important role. The execution of the physical plan may require the cooperation of various peers, and leads to results being sent at the query peer (step~4).

%% file: patterns.tex
\Section{Views, queries and rewritings}
\label{sec:rew-outline}

Once an ad-hoc query is issued by a peer, as described in Section~\ref{sec:platformOverview}, a DHT lookup retrieves the definitions of the existing ViP2P network views which could be used to answer the query (for more details, see Section~\ref{sec:index}). Then, the query peer runs its own algorithm for rewriting the query using the respective materialized views. The algorithm used in ViP2P is presented in~\cite{rewritingICDE2011} and its details are out of the scope of this paper, where we are mainly concerned with the platform and its scalability. However, to make this paper self-contained, we present the XQuery dialect we consider (Section~\ref{sec:xq-dialect}), we present a joined tree pattern formalism that conveniently represents queries and views (Section~\ref{sec:tp}) and describe our algebraic rewritings based on views (Section~\ref{sec:rw}).

\begin{figure}[t!]
\begin{center}
\scalebox{1}{
\begin{tabular}{|c|rp{0.7\columnwidth}|}
\hline
1&$q\;$:=& \kwd{for} $absVar$ (\kwd{,} ($absVar \vert relVar$))* \\&&(\kwd{where} $pred$ (\kwd{and} $pred$)*)? \kwd{return} $ret$\\
\hline
2&$absVar\;$:= &$x_i$ \kwd{in doc(}$uri$\kwd{)} $p$\\
\hline
3&$relVar\;$:=&$x_i$ \kwd{in} $x_j$ $p$ $\qquad$ {\em // $x_j$ introduced before $x_i$}\\
\hline
4&$pred\;$ := & \kwd{string($x_i$) =} $($\kwd{string($x_j$)} $\vert$ $c$$)$\\
\hline
5&$ret\;$:=&$\langle$$l$$\rangle$ $elem$* $\langle /l\rangle$\\
\hline
6&$elem\;$:=& $\langle l_i\rangle$\kwd{\{}  $(x_k \;\vert\;$
\kwd{id($x_k$)} $\;\vert\; $ \kwd{string($x_k$)$)$ \}}$\langle /l_i\rangle$\\
\hline
\end{tabular}
}
\end{center}
\vspace{-5mm}
\caption{Grammar for views and queries.\label{tab:xq}}
\vspace{-4.5mm}
\end{figure}

\Subsection{XQuery dialect}
\label{sec:xq-dialect}

Let ${\cal L}$ be a set of  XML node names, and ${\cal XP}$ be the XPath$^{\{/,//,[\,]\}}$ language~\cite{Suciu}. We consider views and queries expressed in the XQuery dialect described in Figure~\ref{tab:xq}. In the \kwd{for} clause, $absVar$ corresponds to an absolute variable declaration, which binds a variable named $x_i$ to a path expression $p\in {\cal XP}$ to be evaluated starting from the root of some document available at the URI $uri$. The non-terminal $relVar$ allows binding a variable named $x_i$ to a path expression $p\in{\cal XP}$ to be evaluated starting from the bindings of a previously-introduced variable $x_j$. The optional \kwd{where} clause is a conjunction over a number of predicates, each of which compares the string value of a variable $x_i$, either with the string value of another variable $x_j$,  or with a constant $c$.

The \kwd{return} clause builds, for each tuple of bindings of the \kwd{for} variables, a new element labeled $l$, having some children labeled $l_i$ ($l,l_i\in {\cal L}$). Within each such child, we allow one out of three possible information items related to the current binding of a variable $x_k$, declared in the \kwd{for} clause: (1)~$x_k$ denotes the full subtree rooted at the binding of $x_k$;
(2)~\kwd{string($x_k$)} is the string value of the binding; (3)~\kwd{id($x_k$)} denotes the ID of the node to which $x_k$ is bound.

There are important differences between the {\em subtree} rooted at an element (or, equivalently, its {\em content}), its {\em string value} and its {\em ID}. The content of $x_i$ includes all (element, attribute, or text)  descendants  of $x_i$, whereas the string value is only a concatenation of $n$'s text descendants~\cite{XPFN}. Therefore, \kwd{string($x_i$)} is very likely smaller than $x_i$'s content, but it holds less information. Second, an XML ID does not encapsulate the content of the corresponding node. However, XML IDs enable joins which may stitch together tree patterns into larger ones. {\em We assume structural IDs}, i.e., comparing the IDs of two nodes $n_1$ and $n_2$ allows determining if $n_1$ is a parent (or ancestor) of $n_2$. Our XQuery dialect distinguishes structural IDs, value and contents, and allows any subset of the three to be returned for any of the variables, resulting in significant flexibility.

\noindent\begin{figure}[t!]
\begin{center}
\scalebox{0.9}{
\begin{tabular}{||@{}l|l@{}||}
\hline\hline
&\textsf{\small for $\quad\,$ \$p in doc("confs")//confs//SIGMOD/paper, \$y1 in \$p/year,}\\
&\textsf{\small \phantom{return} \$a in \$p//author[email], \$c1 in \$a/affiliation//country,}\\
$\;q$&\textsf{\small \phantom{return} \$b in doc("books")//book, \$y2 in \$b/year, \$e in \$b/editor,}\\
&\textsf{\small \phantom{return} \$t in \$b//title, \$c2 in \$b//country}\\
&\textsf{\small where \$e=`ACM' and \$y1=\$y2 and \$c1=\$c2}\\
&\textsf{\small return $\langle$res$\rangle$ $\langle$tval$\rangle$\{string(\$t)\}$\langle$/tval$\rangle$ $\langle$/res$\rangle$}\\
%
\hline
$\;v_1$&\textsf{\small for $\quad\,$ \$p in doc("confs")//confs//paper, \$a in \$p/affiliation}\\
&\textsf{\small return $\langle$v1$\rangle$ $\langle$pid$\rangle$\{id(\$p)\}$\langle$/pid$\rangle$ $\langle$aid$\rangle$\{id(\$a)\}$\langle$/aid$\rangle$}\\
&\textsf{\small \phantom{return $\langle$v2$\rangle$} $\langle$acont$\rangle$\{\$a\}$\langle$/acont$\rangle$ $\langle$/v1$\rangle$}\\
\hline
&\textsf{\small for $\quad\,$ \$b in doc("books")//book, \$c in \$b//country, \$e in \$b/editor,}\\
&\textsf{\small \phantom{return} \$t in \$b/title, \$y1 in \$b/year, \$p in doc("confs")//SIGMOD/paper,$\;$}\\
$\;v_2$&\textsf{\small \phantom{return} \$y2 in \$p/year, \$a in \$p//author[email]}\\
&\textsf{\small where \$e=`ACM' and \$y1=\$y2}\\
&\textsf{\small return $\langle$v2$\rangle$ $\langle$cval$\rangle$\{string(\$c)\}$\langle$/cval$\rangle$ $\langle$tval$\rangle$\{string(\$t)\}$\langle$/tval$\rangle$}\\
&\textsf{\small \phantom{return $\langle$v2$\rangle$} $\langle$pid$\rangle$\{id(\$p)\}$\langle$/pid$\rangle$ $\langle$aid$\rangle$\{id(\$a)\}$\langle$/aid$\rangle$ $\langle$/v2$\rangle$}\\
\hline
&\textsf{\small for $\quad\,$ \$v1 in doc("v1.xml")//v1, \$p1 in \$v1/pid, \$af1 in \$v1/aid,}\\
&\textsf{\small \phantom{return} \$c1 in \$v1//acont//country, \$v2 in doc("v2.xml")//v2,}\\
$\;r$&\textsf{\small \phantom{return}  \$c2 in \$v2/cval, \$t2 in \$v2/tval, \$p2 in \$v2/pid, \$a2 in \$v2/aid$\;$}\\
&\textsf{\small where \$p1=\$p2 and parent(\$a2,\$af1) and \$c1=\$c2}\\
&\textsf{\small return $\langle$res$\rangle$ $\langle$tval$\rangle$\{\$v2\}$\langle$/tval$\rangle$ $\langle$/res$\rangle$\,}\\
\hline\hline
\end{tabular}
}
\end{center}
\vspace{-4mm}
\caption{XQuery query, views, and rewriting.\label{fig:sample-qv}}
\vspace{-3mm}
\end{figure}

\vspace{-4mm}
For illustration, Figure~\ref{fig:sample-qv} shows a query $q$ in our XQuery dialect, as well as two views $v_1$ and $v_2$. The \textsf{\small parent} custom function returns true if and only if its inputs are node IDs, such that the first identifies the parent of the second. Moreover, as usual in XQuery, the variable bindings that appear in the \textsf{\small where} clauses imply the string values of these bindings (e.g. \$e=`ACM' is implicitly converted to \textsf{\small string}(\$e)=`ACM').

\Subsection{Joined tree patterns}
\label{sec:tp}

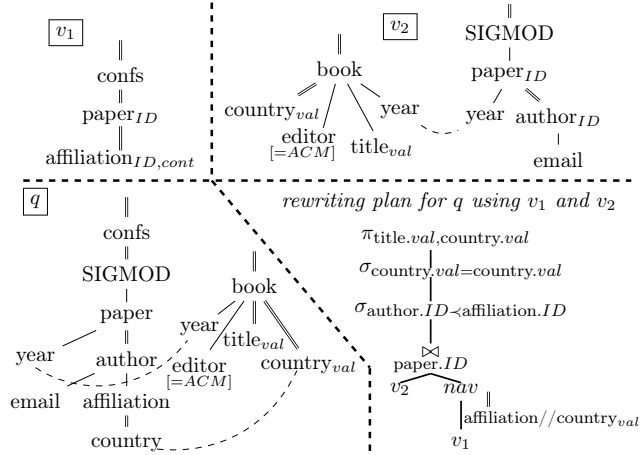
\begin{figure}[t!]
\begin{center}

\scalebox{0.8}{
	\input{fig-queries-views-rewriting.tex}
}
\vspace{-3mm}
\caption{Pattern query and views, and algebraic rewriting.\label{fig:sample-qv-tp}}
\end{center}
\vspace{-6mm}
\end{figure}

We use a dialect of joined tree patterns to represent views and queries. Formally, a tree pattern is a tree whose nodes carry labels from ${\cal L}$ and may be annotated with zero or more among: $ID$, $val$ and $cont$. A pattern node may also be annotated with a value equality predicate of the form $[$=$c]$ where $c$ is some constant. The pattern edges are either simple for parent-child or double for ancestor-descendant relationships. A joined tree pattern is a set of tree patterns, connected through value joins, which are denoted by dashed edges. For illustration, Figure~\ref{fig:sample-qv-tp} depicts the (joined) tree pattern representations of the query and views shown in XQuery syntax in Figure~\ref{fig:sample-qv}. In short, the semantics of an annotated tree pattern against a database is a list of tuples storing the $ID$, $val$ and $cont$ from the tuples of database nodes in which the tree pattern embeds. The tuple order follows the order of the embedding target nodes in the database. The detailed semantics feature some duplicate elimination and projection operators (from the algebra we will detail next), in order to be as close to the W3C's XPath $2.0$ semantics as possible. The only remaining difference is that tree patterns return tuples, whereas standard XPath/XQuery semantics uses node lists. Algebraic operators for translating between the two are by now well understood~\cite{DBLP:reference/db/ManolescuPV09}.
The semantics of a joined tree pattern is the join of the semantics of its component tree patterns.

Translating from our XQuery dialect to the joined tree patterns is quite straightforward. The only part of the XQuery syntax {\em not} reflected in the joined tree patterns is the names of the elements created by the \kwd{return} clause. These names are not needed when rewriting queries based on views. Once a rewriting has been found, the query execution engine creates new  elements out of the returned  tuples of XML elements, values and/or identifiers, using the names specified by the original query, as explained in~\cite{JaiOrdered}. From now on, for readability, we will only use the tree pattern query representations of views and queries.

\Subsection{Rewritings \& algebra}
\label{sec:rw}

A rewriting  is an XQuery query expressed in the same dialect as our views and queries, but formulated against XML documents corresponding to materialized views. For instance, the rewriting XQuery expression $r$ in Figure~\ref{fig:sample-qv} is an equivalent rewriting of the query $q$ using the views $v_1$ and $v_2$ in the same Figure.

An alternative, more convenient, way to view rewritings is under the form of {\em logical algebraic plans}. We will now present the logical operators that are used to express the view rewritings. We denote by $\prec$ the {\em parent comparison operator}, which takes as input arguments two IDs and returns true if the node corresponding to the left-hand ID argument is the parent of the node corresponding to the right-hand ID.
The {\em ancestor comparison operator}, denoted $\anc$, is defined in a similar way. Observe that $\prec$ and $\anc$ are only abstract operators here (we do not make any assumption on how they are evaluated).

We consider an algebra on tuple collections (such as described in the previous Section)  whose main operators are:  (1)~scan of all tuples from a view $v$, denoted $scan(v)$ (or simply $v$ for brevity, whenever possible), (2)~cartesian product, denoted $\times$; (3)~selection, denoted $\sigma_{pred}$, where $pred$ is a conjunction of predicates of the form $a \odot \underline{c}$ or $a\odot b$, $a$ and $b$ are tuple attributes, $\underline{c}$ is some constant, and $\odot$ is a binary operator among $\{=,\prec,\precc\}$; (4)~projection, denoted $\pi_{cols}$, where $cols$ is the attributes list that will be projected; (5)~navigation, denoted $nav_{a,np}$, which is a unary algebraic operator, parameterized by one of its input columns' name $a$, and a tree pattern $np$. Column $a$ must correspond to a $cont$ attribute in the input of $nav$. Let $t$ be a tuple in the input of $nav$, and $np(t.a)$ be the result of evaluating the pattern $np$ on the XML fragment stored in $t.a$. Then, $nav_{a,np}$ outputs the tuples $\{t \times np(t.a)\}$, for each tuple $t$ of the input.

Figure~\ref{fig:samplenav} illustrates the effect of $nav$ when applied on a sample input operator $op$. The parameters to this $nav$ are $cont_{book}$, the name of the column containing $\langle$\textsf{\small book}$\rangle$ elements, and the tree pattern $//author$.  The first tuple output by $nav$ is obtained by augmenting the corresponding input tuple with a $cont_{author}$ attribute containing the single \textsf{\small author}-labeled child of the element found in its $cont_{book}$ attribute. The second and third $nav$ output tuples are similarly obtained from the last tuple produced by $op$. Observe that the second tuple in $op$'s output has been eliminated by the $nav$ since it had no $\langle$\textsf{\small author}$\rangle$ element in its $cont_{book}$ attribute.

The algebra also includes the join operator, defined as usual, sort and duplicate elimination. For illustration, in the bottom of Figure~\ref{fig:sample-qv-tp}, we depict the algebraic representation of the rewriting $r$ shown in XQuery syntax at the bottom of Figure~\ref{fig:sample-qv}.

\begin{figure*}[t!]
\centering
\begin{tabular}{@{}ccc@{}}
$op$ & & $nav_{cont_{book},//author}(op)$\\[3mm]
\scalebox{0.7}{\begin{tabular}{||c|l||}
\hline\hline
$ID_{book}$ & $cont_{book}$\\
\hline\hline
$id_{book,1}$ & \textsf{\small $\langle$book$\rangle$} \\ & \textsf{~~\small $\langle$author$\rangle$author1$\langle$/author$\rangle$} \\ & \textsf{\small $\langle$/book$\rangle$}\\
\hline
$id_{book,2}$ & \textsf{\small $\langle$author/$\rangle$}\\
\hline
$id_{book,3}$ & \textsf{\small $\langle$book$\rangle$}  \\ & \textsf{~~\small  $\langle$author$\rangle$author2$\langle$/author$\rangle$} \\ & \textsf{~~\small $\langle$author$\rangle$author3$\langle$/author$\rangle$} \\ & \textsf{\small $\langle$/book$\rangle$}\\	 
\hline\hline
\end{tabular}}&
$\Rightarrow$ &
\scalebox{0.7}{\begin{tabular}{||c|l|l||}
\hline\hline
$ID_{book}$ & $cont_{book}$ & $cont_{author}$\\
\hline\hline
$id_{book,1}$ & \textsf{\small $\langle$book$\rangle$} & \textsf{\small $\langle$author$\rangle$author1$\langle$/author$\rangle$} \\ & \textsf{~~\small$\langle$author$\rangle$author1$\langle$/author$\rangle$}  & \\ & \textsf{\small$\langle$/book$\rangle$} & \\
\hline
$id_{book,3}$ & \textsf{\small $\langle$book$\rangle$} & \textsf{\small $\langle$author$\rangle$author2$\langle$/author$\rangle$} \\ & \textsf{~~\small$\langle$author$\rangle$author2$\langle$/author$\rangle$}  & \\ & \textsf{~~\small$\langle$author$\rangle$author3$\langle$/author$\rangle$} & \\ & \textsf{\small$\langle$/book$\rangle$} & \\
\hline
$id_{book,3}$ & \textsf{\small $\langle$book$\rangle$} & \textsf{\small $\langle$author$\rangle$author3$\langle$/author$\rangle$} \\ & \textsf{~~\small$\langle$author$\rangle$author2$\langle$/author$\rangle$}  & \\ & \textsf{~~\small$\langle$author$\rangle$author3$\langle$/author$\rangle$} &  \\ & \textsf{\small$\langle$/book$\rangle$} & \\
\hline\hline
\end{tabular}}
 \\
\end{tabular}
\vspace{-1mm}
\caption{Sample input and output to a logical $nav$ operator.
\label{fig:samplenav}}
\vspace{-3mm}
\end{figure*}

An important feature of the rewritings we consider is \emph{minimality}: our rewriting algorithm~\cite{rewritingICDE2011} outputs only rewriting expressions in which no view instance can be removed and still get an equivalent rewriting for a given query.
For instance, the rewriting plan in Figure~\ref{fig:sample-qv-tp}, of the form $\pi(\sigma(v_1\bowtie v_2))$, is minimal. In contrast, a rewriting for the same query of the form $\pi(\sigma(v_1 \bowtie v_2 \bowtie v_3))$, using also view $v_3$ although it is not needed, is not minimal. Considering only minimal rewritings allows for more efficient query execution plans: a non-minimal plan is always less efficient in terms of query execution time than its corresponding minimal one.

%% file: fig-queries-views-rewriting.tex
\begin{tikzpicture}
\tikzstyle{separation lines}=[dashed, line width=1.2pt]
\tikzstyle{separation lines2}=[line width=1.2pt]
\draw[style=separation lines] (-1.7,7.55) -- (8.3,7.55);
\draw[style=separation lines] (1.4,10.5) -- (1.4,7.55);
\draw[style=separation lines] (1.4,7.55) -- (4,4.45);
\draw[style=separation lines] (4,4.44) -- (4.0,3);


\def\vtwoxoffset{0.2}

\draw(-1,10) node (v1) {\fbox{$v_1$}};
\draw(-0.1,10) node (top3) {~};
\draw(-0.1,9.3) node (confs3) {confs};
\draw(-0.1,8.6) node (paper3) {paper$_{ID}$};
\draw(-0.1,7.9) node (affil3) {affiliation$_{ID,cont}$};
\draw(top3) edge[double] (confs3);
\draw(confs3) edge[double] (paper3);
\draw(-.1,8.45) edge[double] (-0.1,8.05);
\draw(4.3+\vtwoxoffset,10.1) node (v2) {\fbox{$v_2$}};
\draw(3.3+\vtwoxoffset,10.1) node (top1) {~};
\draw(3.3+\vtwoxoffset,9.4) node (book1) {book};
\draw(2.2+\vtwoxoffset,8.7) node (country1) {country$_{val}$};
\draw(2.8+\vtwoxoffset,8.3) node (editor1) {editor};
\draw(3.0+\vtwoxoffset,8.3) node (editorInv1) {};
\draw(2.7+\vtwoxoffset,8) node (editorVal1) {$_{[=ACM]}$};
\draw(4.0+\vtwoxoffset,8.1) node (title1) {title$_{val}$};
\draw(4.3+\vtwoxoffset,8.7) node (year1) {year};
\draw(top1) edge[double] (book1);
\draw(book1) edge[double] (country1);
\draw(book1) edge (editorInv1);
\draw(book1) edge (title1);
\draw(book1) edge (year1);
\draw(6.1+\vtwoxoffset,10.6) node (top2) {~};
\draw(6.1+\vtwoxoffset,10) node (ICDE2) {SIGMOD};
\draw(6.1+\vtwoxoffset,9.3) node (paper2) {paper$_{ID}$};
\draw(5.7+\vtwoxoffset,8.6) node (year2) {year};
\draw(6.9+\vtwoxoffset,8.6) node (author2) {author$_{ID}$};
\draw(6.9+\vtwoxoffset,7.9) node (email2) {email};
\draw(top2) edge[double] (ICDE2);
\draw(ICDE2) edge (paper2);
\draw(paper2) edge (year2);
\draw(paper2) edge[double] (author2);
\draw(year1) edge[dashed, bend right=30] (year2);
\draw(author2) edge (email2);
\draw(-1.5,7.2) node (q) {\fbox{$q$}};
\draw(0,7.4) node (top4) {~};
\draw(0,6.7) node (confs4) {confs};
\draw(0,6.0) node (ICDE4) {SIGMOD};
\draw(0,5.3) node (paper4) {paper};
\draw(-1.5,4.6) node (year4) {year};
\draw(0,4.6) node (author4) {author};
\draw(0,3.9) node (affil4) {affiliation};
\draw(-1.5,3.9) node (email4) {email};
\draw(0,3.2) node (country4) {country};
\draw(top4) edge[double] (confs4);
\draw(confs4) edge[double] (ICDE4);
\draw(ICDE4) edge (paper4);
\draw(paper4) edge[double] (author4);
\draw(author4) edge (affil4);
\draw(author4) edge (email4);
\draw(affil4) edge[double] (country4);
\draw(paper4) edge (year4);
\draw(2.1,6.5) node (top5) {~};
\draw(2.1,5.8) node (book5) {book};
\draw(1.2,5.1) node (year5) {year};
\draw(1.2,4.5) node (editor5) {editor};
\draw(1.2,4.2) node (editorVal5) {$_{[=ACM]}$};
\draw(2.1,4.9) node (title5) {title$_{val}$};
\draw(3,4.5) node (country5) {country$_{val}$};
\draw(top5) edge[double] (book5);
\draw(book5) edge (year5);
\draw(book5) edge (editor5);
\draw(book5) edge[double] (title5);
\draw(book5) edge[double] (country5);
\draw(-1.5,4.45) edge[dashed, bend right=43] (1,4.9);
\draw(0.5,3.1) edge[dashed, bend right=35] (2.8,4.4);

\def\yoffsettree{4.5}

\draw(5.3,7.2) node (plan3) {\emph{rewriting plan for $q$ using $v_1$ and $v_2$}};
\draw(5.25,2.1+\yoffsettree) node (project) {$\pi_{\mbox{\footnotesize{title.$val$}},{\mbox{\footnotesize{country.$val$}}}}$};

\draw(5.5,1.55+\yoffsettree) node (sigma2) {$\sigma_{\mbox{\footnotesize{country.$val$}}={\mbox{\footnotesize{country.$val$}}}}$};
\draw(5.5,0.9+\yoffsettree) node (sigma1) {$\sigma_{\mbox{\footnotesize{author.$ID$}}\prec{\mbox{\footnotesize{affiliation.$ID$}}}}$};
\def\yoffsettree{4.3}
\draw(5,.4+\yoffsettree) node (join3) {$\bowtie$};
\draw(5,.2+\yoffsettree) node (jcond3) {\footnotesize{paper.$ID$}};
\draw(5.5,-.2+\yoffsettree) node (navj) {$nav$};
\draw(5.5,-1.1+\yoffsettree) node (v23) {$v_1$};

\draw(5.95,-.25+\yoffsettree) edge[double] (5.95, -.5+\yoffsettree);
\draw(7,-.7+\yoffsettree) node (countrysmall2) {\footnotesize{affiliation//country$_{val}$}};
\draw(4.5,-.2+\yoffsettree) node(v13) {$v_2$};
\draw(navj) edge[thick] (v23);

\draw(5,.05+\yoffsettree) edge[thick] (5.5,-.1+\yoffsettree);
\draw(5,.05+\yoffsettree) edge[thick] (4.5,-.1+\yoffsettree);
\def\yoffsettree{4.5}

\draw(5,.35+\yoffsettree) edge[thick] (5,0.75+\yoffsettree);
\draw(5,1.05+\yoffsettree) edge[thick] (5,1.45+\yoffsettree);

\draw(5,1.9+\yoffsettree) edge[thick] (5,1.6+\yoffsettree);

\end{tikzpicture}

%% file: internals.tex
\Section{ViP2P view management}
\label{sec:viewIndexingAndRetrieval}

Materialized views stand at the heart of data sharing in ViP2P. Sections~\ref{sec:digestAndRepository} and \ref{sec:index} show how view definitions are indexed and looked up in the DHT in order to be retrieved for view materialization and query rewriting, respectively.

\Subsection{View definition indexing \& lookup for view materialization}
\label{sec:digestAndRepository}

This Section describes how published data and views ``meet'', i.e., how ViP2P ensures that for each view $v$, the data obtained by evaluating $v$ over $d$, denoted $v(d)$, is {\em eventually} computed and stored at the peer having defined $v$. Two cases arise, depending on the publication order of $v$ and $d$. 

\vspace{1mm}
\noindent\textbf{View published before the document} In this case, the view definitions are indexed using as keys all the labels (node names and words) of the view. Figure~\ref{fig:viewsAndQueries} shows eight views. To index $v_1$, ViP2P issues the calls $put(book, v_1)$ and $put(title, v_1)$ to the DHT. Observe that these calls index the \emph{definition} of $v_1$ (\emph{not} its data) on the keys $book$ and $title$. Similarly, $v_2$ is indexed on the keys $book$, $author$ and $last$, $v_3$ using the keys $paper$, $author$ and $last$ etc. When the document in Figure~\ref{fig:xmlDoc} is published, {\em get} calls are issued with the keys $bibliography$, $book$, $paper$, $title$, $author$, $year$, $Found.$ $of$ $Databases$ and all the other labels and keywords of the document. The result is a superset of view definitions of the views that the document might affect. In this case the views $v_1$ to $v_8$ are retrieved.  

\begin{figure}[h!]
	\begin{center}
		\vspace{-2mm}
		
		\scalebox{1}{
		\begin{tikzpicture}
			\draw (0cm, 0cm) -- (5cm, 0cm);
			\foreach \x in {0, 1, 2, 3, 4, 5} \draw (\x cm, 3pt) -- (\x cm, - 3pt);
			\draw (0cm, 0cm) node[below=5pt] {$t_i$};
			\draw (1cm, 0cm) node[below=5pt] {$t_{i+1}$};
			\draw (2cm, 0cm) node[below=5pt] {$t_{i+2}$};
			\draw (3cm, 0cm) node[below=5pt] {$t_{i+3}$};
			\draw (4cm, 0cm) node[below=5pt] {$t_{i+4}$};
			\draw (5cm, 0cm) node[below=5pt] {$t_{i+5}$};
			\fill (0.5cm, 0cm) circle (1pt);\draw (0.5cm, 0cm) node[above=3pt] {$d_1$};
			\fill (1.5cm, 0cm) circle (1pt);\draw (1.5cm, 0cm) node[above=3pt] {$v_1$};
			\fill (2.5cm, 0cm) circle (1pt);\draw (2.5cm, 0cm) node[above=3pt] {$v_2$};
			\fill (3.5cm, 0cm) circle (1pt);\draw (3.5cm, 0cm) node[above=3pt] {$v_3$};
			\fill (4.5cm, 0cm) circle (1pt);\draw (4.5cm, 0cm) node[above=3pt] {$d_2$};
		\end{tikzpicture}
		}
		\vspace{-2mm}
		\caption{Sample timeline of view and document publication.\label{fig:timeintervals}}
		\vspace{-3mm}
\end{center}
\end{figure}
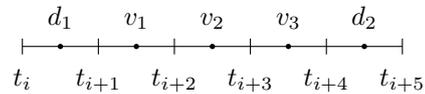

\vspace{1mm}
\noindent\textbf{View published after the document} ViP2P ensures that views are kept up to date (providing some time for the data to circulate across the network). Thus, when a view is published, it should be filled in with data from all the previously published documents matching the view. To achieve this, ViP2P associates to each view an {\em interval timestamp}, corresponding to a time interval during which the view was published, and indexes each view definition in the DHT using as key the corresponding timestamp. As illustrated in Figure~\ref{fig:timeintervals}, $v_1$ belongs to (was published in) the interval $(t_{i+1},t_{i+2}]$, $v_2$ to the interval $(t_{i+2},t_{i+3}]$ and $v_3$ to $(t_{i+3},t_{i+4}]$.

\begin{figure}[t!]
\begin{center}
\scalebox{0.67}{\begin{tabular}{||c||}
\hline\hline
\begin{tabular}{p{1.5cm}|p{1.5cm}|p{1.5cm}|p{1.5cm}|p{2.7cm}|p{2.7cm}|p{2.7cm}}
\begin{tikzpicture}
[level/.style={level distance=10mm}]
\tikzstyle{edge from parent}=[opacity=1,thick,draw]
\tikzstyle{level 1}=[sibling distance=3cm,level distance=7mm]
\tikzstyle{level 2}=[sibling distance=1.4cm]
\raisebox{9mm}{
\node[pname] {$v_1$}
	child {node {$book_{ID}$} edge from parent [transparent edge]
		child {node[plain] {$title_{val}$}}
	}
;
}
\end{tikzpicture}
&
\hspace{0.5mm}
\begin{tikzpicture}
[level/.style={level distance=10mm}]
\tikzstyle{edge from parent}=[opacity=1,draw]
\tikzstyle{level 1}=[sibling distance=3cm,level distance=7mm]
\tikzstyle{level 2}=[sibling distance=1.4cm]
\raisebox{-1mm}{
\node[pname] {$v_2$}
	child {node {$book_{ID}$} edge from parent [transparent edge]
		child {node[plain] {$author$}
			child {node[plain] {$last_{val}$} edge from parent [descendant]}
		}
	}
;
}
\end{tikzpicture}
&
\begin{tikzpicture}
[level/.style={level distance=10mm}]
\tikzstyle{edge from parent}=[opacity=1,draw]
\tikzstyle{level 1}=[sibling distance=3cm,level distance=7mm]
\tikzstyle{level 2}=[sibling distance=1.4cm]
\raisebox{-1mm}{
\node[pname] {$v_3$}
	child {node {$paper_{ID}$} edge from parent [transparent edge]
		child {node[plain] {$author$}
			child {node[plain] {$last_{val}$} edge from parent [descendant]} 
		}
	}
;
}
\end{tikzpicture}
&
\begin{tikzpicture}
[level/.style={level distance=10mm}]
\tikzstyle{edge from parent}=[opacity=1,draw]
\tikzstyle{level 1}=[sibling distance=3cm,level distance=7mm]
\tikzstyle{level 2}=[sibling distance=1.4cm]
\raisebox{-1mm}{
\node[pname] {$v_4$}
	child {node {$paper_{ID}$} edge from parent [transparent edge]
		child {node[plain] {$author$}
			child {node[plain] {$first_{val}$} edge from parent [descendant]}
		}
	}
;
}
\end{tikzpicture}
&
\begin{tikzpicture}
[level/.style={level distance=10mm}]
\tikzstyle{edge from parent}=[opacity=1,draw]
\tikzstyle{level 1}=[sibling distance=3cm,level distance=7mm]
\tikzstyle{level 2}=[sibling distance=1.4cm]
\raisebox{9mm}{
\node[pname] {$v_5$}
	child {node {$book$} edge from parent [transparent edge]
		child {node[plain] {$title_{val}$}}
		child {node[plain] {$author$}}
	}
;
}
\end{tikzpicture}
&
\begin{tikzpicture}
[level/.style={level distance=10mm}]
\tikzstyle{edge from parent}=[opacity=1,draw]
\tikzstyle{level 1}=[sibling distance=3cm,level distance=7mm]
\tikzstyle{level 2}=[sibling distance=1.4cm]
\raisebox{9mm}{
\node[pname] {$v_6$}
	child {node {$book$} edge from parent [transparent edge]
		child {node[plain] {$title_{val}$}}
		child {node[plain] {$author_{val}$}}
	}
;
}
\end{tikzpicture}
&
\begin{tikzpicture}
[level/.style={level distance=10mm}]
\tikzstyle{edge from parent}=[opacity=1,draw]
\tikzstyle{level 1}=[sibling distance=3cm,level distance=7mm]
\tikzstyle{level 2}=[sibling distance=1.4cm]
\raisebox{9mm}{
\node[pname] {$v_7$}
	child {node {$paper$} edge from parent [transparent edge]
		child {node[plain] {$author_{val}$}}
		child {node[plain] {$year_{val}$}}
	}
;
}
\end{tikzpicture}
\end{tabular}
\\
\hline
\begin{tabular}{c|c|c|c}
\begin{tikzpicture}
[level/.style={level distance=10mm}]
\tikzstyle{edge from parent}=[opacity=1,draw]
\tikzstyle{level 1}=[sibling distance=2.3cm,level distance=7mm]
\tikzstyle{level 2}=[sibling distance=1.4cm]
\raisebox{7mm}{
\node[pname] {$v_8$}
	child {node {$book_{ID}$} edge from parent [transparent edge]
		child {node[plain] (a1){$author$}}
	}
	child {node {paper} edge from parent [transparent edge]
		child {node[plain] (a2){$author$}} 
		child {node[plain] {$year_{val}$}}
	}
;
\draw(a1) edge[dashed, bend right=30] (a2);
}
\end{tikzpicture}
&
\begin{tikzpicture}
[level/.style={level distance=10mm}]
\tikzstyle{edge from parent}=[opacity=1,draw]
\tikzstyle{level 1}=[sibling distance=3cm,level distance=7mm]
\tikzstyle{level 2}=[sibling distance=1.4cm]
\raisebox{-1mm}{
\node[pname] {$q_1$}
	child {node {$book$} edge from parent [transparent edge]
		child {node[plain] {$title_{val}$}} 
		child {node[plain] {$author$}
			child {node[plain] {$last_{val}$} edge from parent [descendant] }		
		}
	}
;
}
\end{tikzpicture}
&
\begin{tikzpicture}
[level/.style={level distance=10mm}]
\tikzstyle{edge from parent}=[opacity=1,draw]
\tikzstyle{level 1}=[sibling distance=3cm,level distance=7mm]
\tikzstyle{level 2}=[sibling distance=1.4cm]
\raisebox{9mm}{
\node[pname] {$q_2$}
	child {node {$book$} edge from parent [transparent edge]
		child {node[plain] {$author$}}
		child {node[plain] {$year_{val}$}}
	}
;
}
\end{tikzpicture}
&
\begin{tikzpicture}
[level/.style={level distance=10mm}]
\tikzstyle{edge from parent}=[opacity=1,draw]
\tikzstyle{level 1}=[sibling distance=3cm,level distance=7mm]
\tikzstyle{level 2}=[sibling distance=1.4cm]
\raisebox{-1mm}{
\node[pname] {$q_3$}
	child {node {$book$} edge from parent [transparent edge]
		child {node[plain] {$title_{val}$}} 
		child {node[plain] (a1){$author$}}
	}
	child {node {paper} edge from parent [transparent edge]
		child {node[plain] (a2){$author$}} 
		child {node[plain] {$year$}
			child {node[plain] {\underline{$2008$}}}
		}
	}
;
\draw(a1) edge[dashed, bend right=30] (a2);
}
\end{tikzpicture}
\end{tabular}\\
\hline\hline
\end{tabular}}
\end{center}
\vspace{-5mm}
\caption{Sample views and queries.\label{fig:viewsAndQueries}}
\vspace{-2mm}
\end{figure}

Each peer having published a document $d$ must check the DHT for views that may have appeared after $d$. To that effect, each peer performs regular lookups using as key the time interval that has just finished. This retrieves the definitions of all views published during that interval. The peer then checks, for each of its documents, if the document has already contributed to that view (this information is stored locally at the peer). If this is not the case, the peer checks if that document holds any data for these views and if so, extracts and sends the corresponding data to the view holder. In Figure~\ref{fig:timeintervals}, document $d_1$ arrives during the $(t_i,t_{i+1}]$ time interval. With the help of the timestamped view index, we discover the views $v_1$, $v_2$ and $v_3$ which arrived later. Notice also that document $d_2$ is published after the views and thus is treated according to the first case above.

\Subsection{View definition indexing \& lookup for query rewriting}
\label{sec:index}
View definitions are also indexed in order to find views that may be used to rewrite a given query. In this context, a given algorithm for extracting (key, value) pairs out of a view definition is termed a {\em view indexing strategy}. For each such strategy, a {\em view lookup} method is needed, in order to identify, given a query $q$, (a superset of) the views which could be used to rewrite $q$. Many strategies can be devised. We present four that we have implemented, together with the space complexity of the view indexing strategy, and the number of lookups required by the view lookup method. We also show that these strategies are {\em complete}, i.e., they retrieve at least all the views that could be embedded in $q$ and, thus, could potentially lead to $q$ rewritings. 

\Subsubsection{Label indexing (LI)} 
In this strategy we index $v$ by each $v$ node label (either some element or attribute name, or word). The number of (key, value) pairs thus obtained is in $O(|v|)$, where $|v|$ the number of nodes of the view. 

\vspace{1mm}
\noindent\textbf{View lookup for LI} The lookup is performed by all node labels of $q$. The number of lookups is $\varTheta(|q|)$, where $|q|$ is the number of nodes in the query. Figure~\ref{fig:viewsAndQueries} depicts some sample queries. The LI lookup keys for $q_1$ are $book$, $title$, $author$ and $last$, retrieving all the views of Figure~\ref{fig:viewsAndQueries}. Note that some of these cannot be used to equivalently rewrite $q_1$. For instance, $v_3$ has data about papers, while $q_1$ asks for books. Similarly, LI indexing and lookup for $q_2$ and $q_3$ leads to retrieving all the views. This shows that LI has many false positives.

\vspace{1mm}
\noindent\textbf{LI completeness} If LI is not complete, then there exists a view $v$ that can be used to rewrite a query $q$, and $v$ is not retrieved when searching by all $q$ labels. It has been shown~\cite{DBLP:conf/icde/TangYOCW08} that in order for a view to appear in an equivalent rewriting of a query, there must exist an embedding (homomorphism) from the view into the query, which entails that some node labels must appear in both. If in our case $v$ and $q$ have no common node label, this contradicts the hypothesis that $v$ was useful to rewrite $q$. 

The LI strategy coincides with the view definition indexing for document-driven lookup (described previously). An interesting variant can furthermore be elaborated.

\Subsubsection{Return label indexing (RLI)} 
Here, we index $v$ by the labels of all $v$ nodes which project some attributes (at most $|v|$). For instance, in Figure~\ref{fig:viewsAndQueries}, the index keys for $v_1$ are $book$ and $title$, for $v_2$ they are $book$ and $last$, for $v_3$ $paper$ and $last$ etc. up to $v_8$ which is indexed by RLI on the keys $book$ and $year$.

\vspace{1mm}
\noindent\textbf{View lookup for RLI} The view definition lookup is the same as for LI (look up on all query node labels). In Figure~\ref{fig:viewsAndQueries}, the definitions of $v_1-v_3$, and $v_5-v_8$ will be retrieved for $q_1$. For $q_2$, the definitions of $v_1$, $v_2$, $v_6$, $v_7$ and $v_8$ will be retrieved. A RLI lookup for $q_3$ will retrieve $v_1-v_8$. Observe that RLI lead to less view definitions retrieved than LI. 

\vspace{1mm}
\noindent\textbf{RLI completeness} Suppose that there is a view $v$ which can be used to rewrite a query $q$, yet the definition of $v$ is not retrieved by RLI lookup. This means that either ($i$)~$v$ does not store any attributes or $(ii)$~the labels of $v$ nodes that project an attribute do not appear in $q$. $(i)$ is not possible because a view that participates to a rewriting should store at least an attribute and $(ii)$ is also not possible since it contradicts the existence of an embedding from $v$ to $q$, required for $v$ to be useful in rewriting $q$.


\Subsubsection{Leaf path indexing (LPI)} 
Let $LP(v)$ be the set of all the distinct root-to-leaf label paths of $v$. Here, a path is just the sequence of labels encountered as one goes down from the root to the node, and does not reflect the type of the edges.  We index $v$ using each element of $LP(v)$ as key. The number of (key, value) pairs thus obtained is in $\varTheta(|LP(v)|)$. Going back to Figure~\ref{fig:viewsAndQueries}, $v_1$ is indexed on the key $book.title$, $v_2$ with the key $book.author.last$ etc.
The view $v_8$, composed of two tree patterns, is indexed using the keys $book.author$, $paper.author$ and $paper.year$.

\vspace{1mm}
\noindent\textbf{View lookup for LPI} Let $LP(q)$ be the set of all the distinct root-to-leaf label paths of $q$. Let $SP(q)$ be the set of all non-empty sub-paths of some path from $LP(q)$, i.e., each path from $SP(q)$ is obtained by erasing some labels from a path in $LP(q)$. Use each element in $SP(q)$ as lookup key. For example, $q_1$ of Figure~\ref{fig:viewsAndQueries} LPI lookup uses the keys  $book$.$title$, $book$, $title$, $book$.$author$.$last$, $book$.$author$, $author$.$last$, $book$.$last$, $book$, $author$ and $last$ etc.
Note that LPI lookup for $q_1$ does not retrieve the definitions of the views $v_3$, $v_4$, and $v_7$, which previous strategies retrieved, although they are not useful to rewrite $q_1$.  LPI can still have some false positives though: a lookup for $q_2$ retrieves $v_5$, $v_6$ and $v_8$, none of which can be used to rewrite $q_2$ (in this example, $q_2$ simply has no rewriting). The lookup for $q_3$ retrieved the views $v_1$, $v_5$, $v_6$, $v_7$ and $v_8$. The filtering is very good in this case because among these only $v_5$ can not be used to rewrite $q_3$. 

Let $h(q)$ be the height of $q$ and $l(q)$ be the number of leaves in $q$. The number of LPI lookups is bound by $\Sigma_{p\in LP(q)}2^{|p|}\leq l(q)\times 2^{h(q)}$. If the query $q$ is a join of tree patterns ($tpq$s) then the bound becomes $\Sigma_{tpq \in q}(\Sigma_{p\in LP(tpq)}2^{|p|})$. 

\vspace{1mm}
\noindent\textbf{LPI completeness} is guaranteed by the fact that if a view $v$ can be embedded in the query $q$, then $LP(v) \subseteq SP(q)$.

\Subsubsection{Return path indexing (RPI)}
RPI is the last strategy that we consider. Let $RP(v)$ be the set of all rooted paths in $v$ which end in a node that returns some attribute. Index $v$ using each element of $LP(v)$ as key. The number of (key,value) pairs is also in $\varTheta(|RP(v)|)$. The indexing keys for $v_1$ are $book$ and $book$.$title$, for $v_2$ are $book$ and $book$.$author$.$last$ etc.

\vspace{1mm}
\noindent\textbf{View lookup for RPI} coincides exactly with the lookup for LPI. The lookup of $q_1$ retrieves the definitions of the views $v_1$, $v_2$, $v_5$, $v_6$ and $v_8$, the same as LPI.  For $q_2$, RPI lookup retrieves the definitions of $v_1$, $v_2$, $v_6$ and $v_8$. Observe that unlike LPI, RPI in this situation does not return $v_5$, which indeed is not useful! We end by noting that this increase of precision of RPI over LPI is not guaranteed. For example, an RPI lookup for $q_3$ retrieves the definitions of all views in Figure~\ref{fig:viewsAndQueries}, which is much less precise than LPI.

\vspace{1mm}
\noindent\textbf{RPI completeness} is established in a similar fashion to the LPI case.


%% file: experimentation.tex
\Section{Experimental results}
\label{sec:exp}

In this Section we present a set of experiments studying ViP2P performance. Section~\ref{sec:settings} outlines the experimental setup. ViP2P attempts to speed up query processing by exploiting pre-computed materialized views. This shifts the complexity of extracting and sending interesting data across the network, from query processing to view materialization, to which we devote the most attention in our experiments. Several parameters determine view materialization performance: the distribution of the documents and views in the network, the documents which contribute to each view, the documents and views size etc. Section~\ref{sec-experimentation-microbenchmarks} starts by studying view materialization in the context of a single peer. Then, Section~\ref{sec-exp-materialization} examines view materialization in the large, in widely different network configurations, varying the number and the distribution of publisher and consumer peers. Section~\ref{sec:experimental-view-indexing-retrieval} presents an evaluation of the indexing strategies for query rewriting presented in Section~\ref{sec:index}. Finally, Section~\ref{sec-exp-query} presents experiments that evaluate the performance of the query execution engine.

\Subsection{Experimentation settings}
\label{sec:settings}

\noindent\textbf{Infrastructure setup} We have carried our experiments on the Grid5000 infrastructure (https://www.grid5000.fr), providing computational resources distributed over nine major cities across France. Figure~\ref{fig:grid5k} shows Grid5000 network topology. Sites are interconnected with a 10Gbps network and within each site, nodes are interconnected with (at least) 1Gbps Ethernet network. The hardware of Grid5000 machines varies from dual-core machines (of at least 1.6 GHz clock speed) with 2GBs of RAM to 16-core machines with 32GBs of RAM. We settled for a random and heterogeneous distribution of hardware, in order to be close to real P2P deployment scenarios. However, in some experiments, we deliberately choose sites being very far away from each other, almost being the two opposite ends of the network, to show the scalability of our platform in the most difficult scenarios imagined within the Grid5000 network.

\vspace{1mm}
\noindent\textbf{Data generation} To have fine control over all the parameters impacting our experiments, we have used synthetic data, produced by two existing XML data generators: ToXGene~\cite{toxgene} and MemBeR~\cite{membergen}.

\vspace{1mm}
\noindent\textbf{Experimentation parameters} We summarize the main parameters characterizing our experiments in Table~\ref{tab:parameters}. For each set $S$, we use $|S|$ to denote the size of the set. Thus, $|P|$ is the number of peers in the network etc. Finally, for a document $d$, we use $|d|$ to denote the size of $d$, measured in Megabytes (MBs).

\vspace{1mm}
\noindent\textbf{Evaluation metrics} In our measurements, we use the following metrics to characterize the system performance:
\begin{itemize}

\item \textbf{Materialization time} is the time needed for the network to materialize a set of views populating them with the data extracted by all the documents published in the network. The materialization time starts at the time instance that a peer initiates the first extraction of data and ends at the time that all peers have extracted and shipped the tuples to the appropriate view holders.

\begin{figure}[t!]
\begin{center}
\includegraphics[width=0.6\columnwidth]{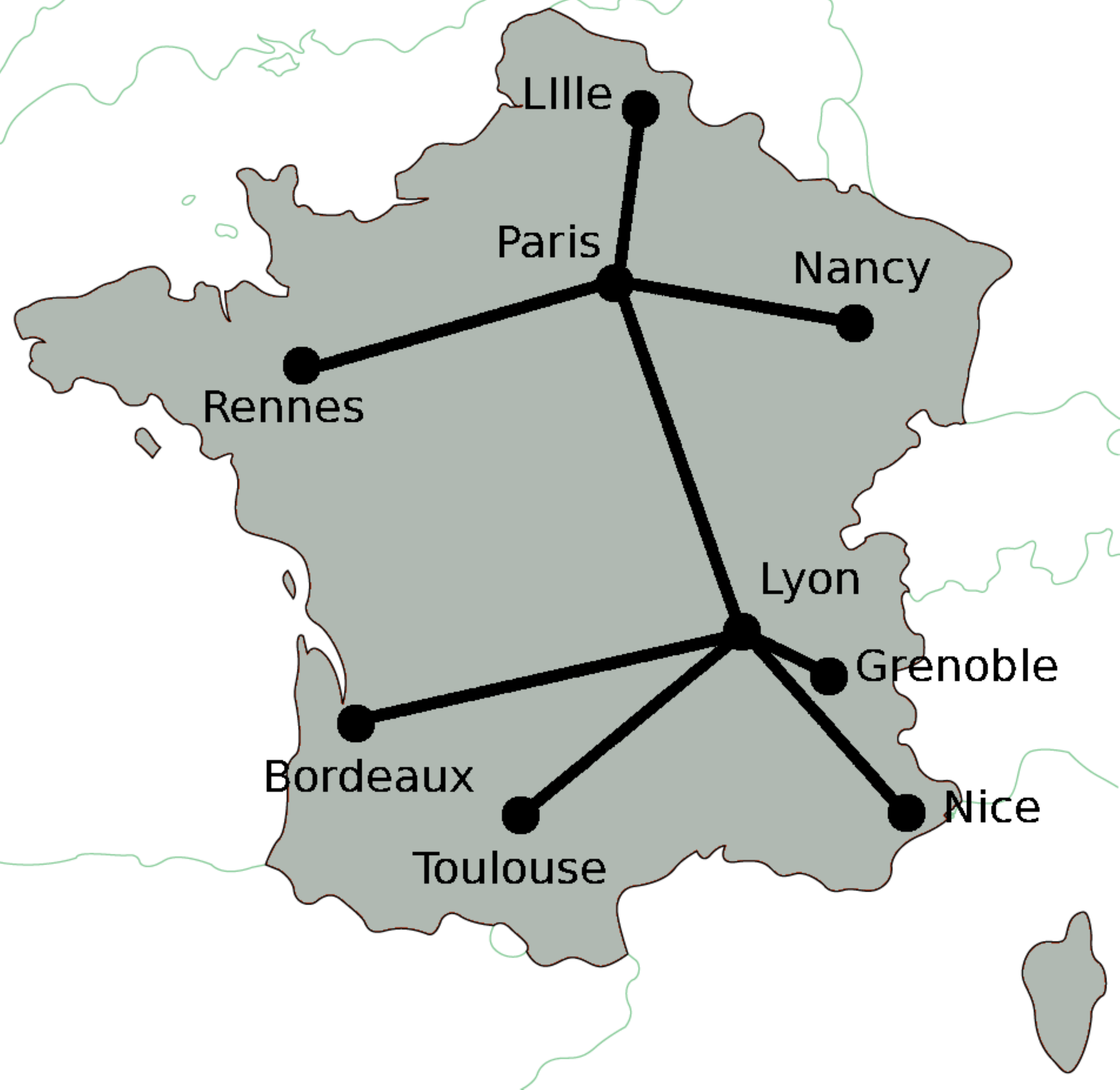}
\caption{Grid5000 network topology.\label{fig:grid5k}}
\end{center}
\end{figure}

\item \textbf{Tuple extraction time} for a view $v$ and a document $d$ is the time needed for the publisher of $d$ to extract from $d$ the tuples which make up $v(d)$.

\item \textbf{Storage time} for a document $d$ and a view $v$ is the time taken by the consumer holding $v$, to add to the corresponding BerkeleyDB database the set of tuples corresponding to $v(d)$.

\item \textbf{Data exchange time} for a document $d$ and view $v$ is the time needed for the tuples $v(d)$ to be transferred across the network from the publisher of $d$ to the consumer holding $v$.

\item \textbf{Lookup time} for a query $q$ is the time needed for the peer asking $q$ to lookup in the DHT the views that may be useful to rewrite $q$.

\item \textbf{Embedding time} for a query $q$ and a set of views $V$ is the time needed by the query peer to verify which of the views may actually be used to rewrite $q$. Recall from Section~\ref{sec:index} that this is established by checking for the presence of embeddings between each view $v\in V$ and the query $q$~\cite{DBLP:conf/icde/TangYOCW08}.

\item \textbf{Query response time} for a query $q$ is the time elapsed between the moment when the query has been posed, and the moment when its execution has finished (as observed at the query peer).

\item \textbf{Time to first result} for a query $q$ is the time between the moment when the query has been posed, and the moment when its first result tuple has been received at the query peer.
\end{itemize}

Whenever the query, view, or document are not specified for a given metric, {\em the metric value is understood to be  the sum, over all the documents, views, and queries used in the respective experiment, of the respective metric, with the exception of the materialization time}. By nature, this metric accounts for many materialization processes running {\em in parallel}, and therefore is not the sum of individual materialization times. For instance, assume publisher $p_1$ publishes a document which contributes data to a view at $p_2$, while publisher $p_1'$ similarly contributes to a view at $p_2'$. The peers $p_1$ and $p_1'$ will start at about the same time the materialization process by looking up views to which they could contribute etc. One of them will be the last to report that all its tuples have been stored and acknowledged by the respective consumer peer. The materialization time of this experiment spans between the first materialization start event, and the last materialization end event, while the two processes run in parallel. 

\begin{table}[t!]
\begin{center}
  \begin{tabular}{ | p{0.12\columnwidth} | p{0.75\columnwidth} | }
\hline
    Symbol    & Description\\ \hline    \hline
    $P$     & The set of peers in the network \\\hline
    $P_{D}$ & The set of peers holding at least one document \\\hline
	$V$     & The set of views in the network \\\hline
	$P_{V}$ & The set of peers holding at least one view \\\hline
	$D$     & The set of all published documents \\\hline
	$D_{V}$ & The set of documents matching at least one view\\\hline
\hline
  \end{tabular}
\vspace{-1mm}
\caption{Parameters characterizing the experiments.
\label{tab:parameters}}
\end{center}
\end{table}

\Subsection{View materialization in the small}
\label{sec-experimentation-microbenchmarks}

We start by studying the performance of extracting from a document $d$, the tuples corresponding to a view $v$, and sending these $v(d)$ tuples from the peer holding $d$ to the one storing $v$. To focus exactly on the process of extraction, we use very simplistic network settings. View materialization in more complex settings and larger scale will be studied next.

\vspace{1mm}
\noindent\textbf{Experiment 1: sequential vs. parallel extraction of views} As described in Section~\ref{sec:archi}, a ViP2P peer $p$ is capable of simultaneously matching several views $v_1,v_2,$ $\ldots,v_k$ on a given document $d$ residing at $p$. The corresponding tuples $v_1(d)$, $v_2(d)$, $\ldots$, $v_k(d)$ are extracted during a single traversal of the document $d$, instead of $k$ traversals (one for each of the $k$ views). This is important when publishing a document $d$ in case the publisher finds out that many previously defined views could match $d$, and therefore it has to match all of them against $d$. While parallel extraction is faster, it may require more memory, since matches for the various views have to be constructed and kept in memory at the same time.


Our first experiment studies the effect of extracting data for several views in parallel. We use a document $d$ and two distinct sets of views. First, we consider a four-view set of the form $\{//t_{i\;ID}\}$ for $i\in \{1,\ldots,4\}$. Second, we consider a larger set including views of the form $\{//t_{i\;ID}\}$ for $i\in \{1,\ldots,8\}$. The views and $d$ are chosen so that $d$ contributes 130.000 tuples to each published view $v_i$. The parameters characterizing the experiment are as follows:

\vspace{-1mm}
\begin{center}
  \begin{tabular}[h!]{ c p{5mm} c  p{5mm} c p{5mm} c p{5mm} c p{5mm} c p{5mm} c p{5mm} c p{5mm} }
    
    $|P|$ & & $|P_{D}|$ & & $|V|$ & & $|P_{V}|$ & & $|D|$ & & $|D_{V}|$ & & $|d|$ \\ \hline
	2 & & 1 & & \{4, 8\} & & 1 & & 1 & & 1 & & 100MB \\
  \end{tabular}
\end{center}

\begin{figure}[t!]
  \begin{center}
	\includegraphics[width=0.49\textwidth]{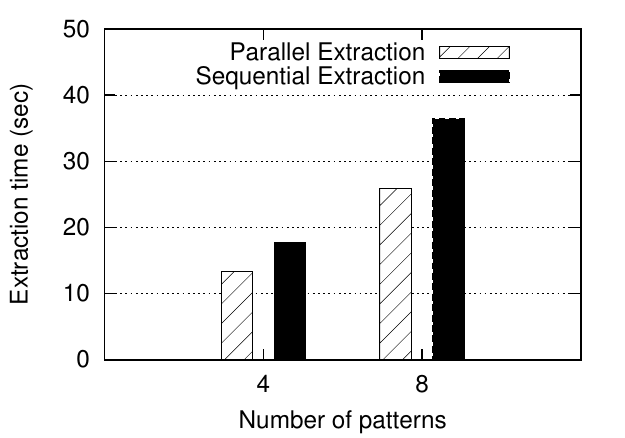}
	\includegraphics[width=0.49\textwidth]{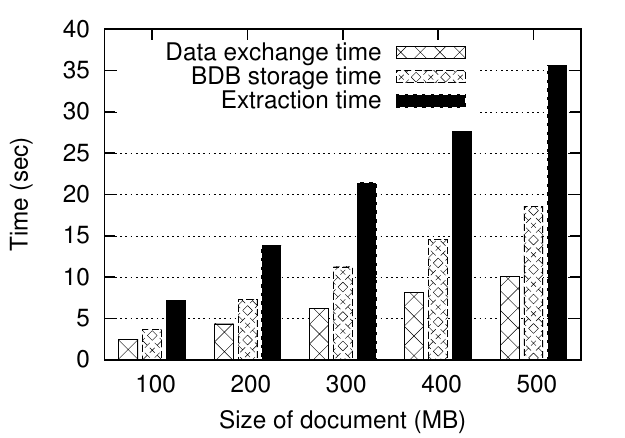}
	\vspace{-2mm}
	\caption{Experiment 1: parallel vs. sequential extraction time (left); experiment 2: view materialization over different-size documents (right). \label{parallel-extraction-data}}
\end{center}
\end{figure}

Figure~\ref{parallel-extraction-data} (left) depicts  the  extraction time when extracting data out of $d$ for four and for eight views, in a parallel and sequential fashion. We observe that parallel extraction accelerates data extraction (in this case, up to 40\%). Therefore,  we will always use parallel extraction in the subsequent experiments. 

\vspace{1mm}
\noindent\textbf{Experiment 2: studying one data transfer pipe} We now study the materialization of documents of various sizes, in order to identify the bottleneck of the materialization process. Possible bottlenecks are $(i)$~data extraction at the document publisher; $(ii)$~network bandwidth between a consumer and a publisher; $(iii)$~view storage time at the consumer. For this experiment, the following parameters are used:

\vspace{-1mm}
\begin{center}
  \begin{tabular}[h!]{ c p{5mm} c  p{5mm} c p{5mm} c p{5mm} c p{5mm} c p{5mm} c p{5mm} c p{5mm} }
    $|P|$ & & $|P_{D}|$ & & $|V|$ & & $|P_{V}|$ & & $|D|$ & & $|D_{V}|$ & & $|d|$ \\ \hline
	2& &1& &1& &1& &1& &1& & \{$100,\ldots,500$\}MB \\
  \end{tabular}
\end{center}

One peer plays the role of the publisher, while the other is the consumer. The peers are located at two opposite ends of France (Lille and Grenoble). The document and the view are chosen so that the complete content of the document is extracted and sent to the consumer, thus, the materialized view size increases linearly to the size of the document. 

Let us now detail the synchronization of the various processes involved when a publisher sends data to a consumer to be added in a view.

\begin{enumerate}
\item The publisher extracts data locally. {\em After} all the tuples from $v(d)$ have been computed, the publisher starts sending them to the consumer\footnote{This could be improved to parallelize extraction and sending in some cases, but there are fundamental limitations: for some of the views we support, one needs to wait for the full traversal of the document before producing an output tuple~\cite{DBLP:conf/lata/GauwinNT09}.}.
\item Packets of tuples are sent over the network to the consumer in an asynchronous way using buffers at the consumer side. 
\item At the consumer, a thread picks packets of tuples from the buffer and stores them in the BerkeleyDB database. 
\end{enumerate}

The buffer at the consumer can be parameterized to control the data transfer speed: when the buffer is full because the storage thread is not sufficiently fast, data transfer stalls.  For this experiment, the size of the data buffer was set to \emph{unlimited} (making sure in advance that the memory of the consumer is enough to store all the produced tuples), so that the data exchange thread can use as much as possible of the available bandwidth between the two peers.

Figure~\ref{parallel-extraction-data} (right) depicts the time needed for the view tuples to be $(i)$~extracted from the document, $(ii)$~sent over the network and $(iii)$~stored in BerkeleyDB at the consumer. We observe that the three times increase linearly in the size of the data. Data extraction is the slowest component, however, overall, times were comparable (also recall that the network connection is fast within the Grid, thus transfer times may be higher in other contexts). 

\vspace{1mm}
\noindent\textbf{Conclusion} From the above two experiments, we conclude that ($i$)~parallelizing data extraction does speed up the time to compute view tuples; ($ii$)~extraction time grows linearly to the size of the input document and ($iii$)~data transfer and data storage time grow linearly with the size of the extracted tuples.

\Subsection{View materialization in larger networks}
\label{sec-exp-materialization}

We now consider view materialization in larger and more complex environments, with many publishers and/or many consumers. 

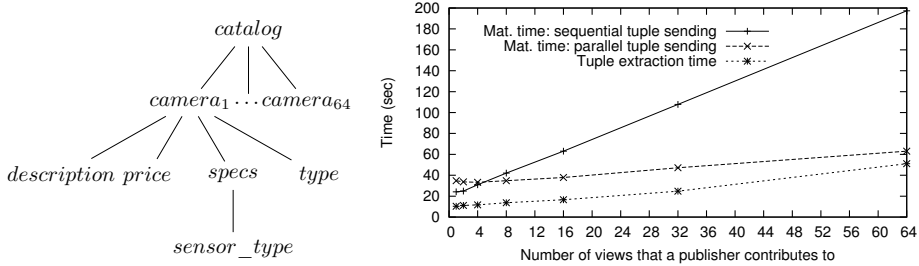
\begin{figure}[t!]
  \begin{center}
	\scalebox{1.01}{
	\scalebox{0.8}{
   	\begin{tikzpicture}[level distance=12mm]
	\tikzstyle{edge from parent}=[draw]
	\tikzstyle{level 1}=[sibling distance=0.95cm]
	\tikzstyle{level 2}=[sibling distance=1.4cm]
	\node {$catalog$}
		child {
			node {$camera_1$}
				child {
					node {$description$}
				}
				child {
					node {$price$}
				}
				child {
					node {$specs$}
					child {
						node {$sensor\_type$}
					}
				}
				child {
					node {$type$}
				}
		}
		child {
			node {\ldots}
		}
		child {
			node {$camera_{64}$}
		};
	\end{tikzpicture}
	}
	\includegraphics[width=0.6\textwidth]{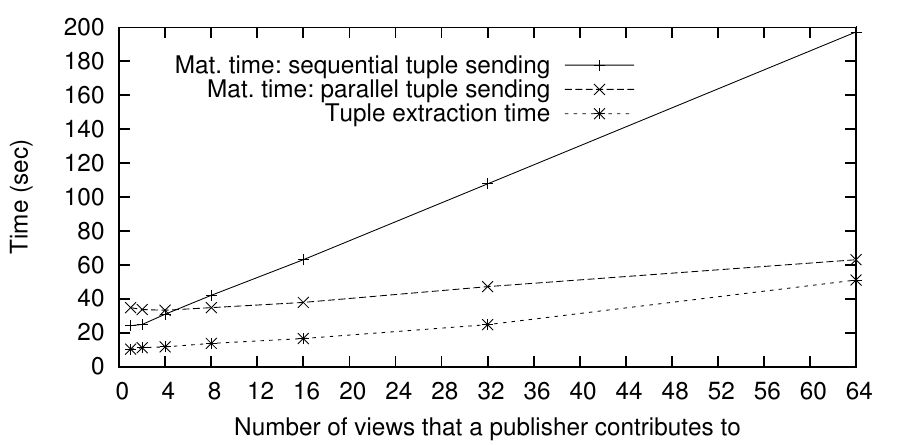}
	}
	\vspace{-6mm}
  \caption{Outline of a controlled synthetic document for our experiments (left); experiment 3: view extraction and materialization time depending on the number of consumers (right).
  \label{fig:64-tag-document_and_views_per_peer}}
\end{center}
\end{figure}

\vspace{1mm}
\noindent\textbf{Documents} For these experiments, we needed to tightly control which parts of the published data are relevant to which views on each peer. Therefore, unless stated otherwise, we rely on documents whose shape is outlined on the left of Figure~\ref{fig:64-tag-document_and_views_per_peer}. There are always 64 camera elements under one $catalog$, and each $camera$ has 4 children. To obtain different document sizes, we insert text of varying length in the $description$ of each $camera$.

\vspace{1mm}
\noindent\textbf{Experiment 3: one publisher, fixed data, varying number of consumers} In this experiment, we use a single publisher, a fixed data set (5 documents of 50 MBs each), and a varying number of consumers (from 1 to 64). Each consumer always holds exactly one view. All the published data is relevant for some view and moreover, the view contents do not overlap, i.e., the data is practically ``partitioned'' over the views. Thus, when there is a single consumer, its view stores the $cont$ of all cameras from the catalog. When there are two consumers, the view of the first consumer stores the $cont$ of the cameras from camera$_1$ to camera$_{32}$, while the other consumer's view stores the $cont$ of the rest of the cameras (camera$_{33}$ to camera$_{64}$) and so on. This way, the views absorb all the data published. The producer is located in Lille and the consumers in Sophia-Antipolis (two opposite ends of France). The parameters values for this experiment are given in the table below:

\vspace{-1mm}
\begin{center}
  \begin{tabular}[h!]{c c c c c c p{1mm} c c}
    $|P|$ & $|P_{D}|$ & $|V|$ & $|P_{V}|$ & $|D|$ & $|D_{V}|$ & &$|d|$ \\\hline
 	65 & 1 & \{1,2,4,\ldots,32,64\} & \{$1,2,4,\ldots,32,64$\} & 5 & 5 & & 50MB \\
  \end{tabular}
\end{center}

Once the tuples are extracted by a publisher, they can be shipped to the view holders sequentially (the publisher contacts the consumers one after the other) or in parallel (the publisher ships all the tuples to all consumers concurrently). At right in Figure~\ref{fig:64-tag-document_and_views_per_peer}, we show the time needed to extract the tuples, and the materialization time for the two variations of tuple sending: sequential or parallel. In both cases, as expected, the extraction time is the same and it increases linearly with the number of consumers.

When sending tuples sequentially, we observe that the materialization time increases linearly with the number of consumers (views). In the case of 64 consumers, data extraction takes about 45 seconds, but materialization takes about 200 seconds. Materialization time increases drastically with sequential tuple sending since more and more consumers need to be contacted one after another.

When sending tuples in parallel, we observe that the materialization time is notably lower than in the case of sequential tuple shipping and that its slope is almost the same as the one of the extraction time. This is because, as soon as the tuples are extracted, a pool of threads (one thread for each packet of tuples) takes over the task of shipping all the tuples in parallel. The bottleneck in this situation is the upload link of each consumer.

\vspace{1mm}
\noindent\textbf{Experiment 4: one publisher, varying data size, 64 consumers}
We study how materialization time is affected when the total size of published data is increased. We use one publisher. The size of the published data varies from 64MBs to 1024MBs.

Each of the 64 consumers holds one view of the form $//catalog//camera_{K\;cont}$ where $K$ varies according to the peer that holds the view. For example, the first consumer holds the view $//catalog//camera_{1\;cont}$, the second holds the view $//catalog//camera_{2\;cont}$ etc. This way, from each document the publisher extracts 64 tuples, each of which is sent to a different consumer.  All the content of the documents is absorbed by the 64 views. The parameter values used for this experiment are: 

\vspace{-1mm}
\begin{center}
  \begin{tabular}[h!]{ c c  c c c c c c p{2mm} }
    $|P|$ & $|P_{D}|$ & $|V|$ & $|P_{V}|$ &  $|D|$ &   $|D_{V}|$     & & $|d|$ \\\hline
 	65    &   1      & 64  &  64     & \{$64,512,1024$\} & \{$64,512,1024$\} & & 1MB   \\
  \end{tabular}
\end{center}

\begin{figure}[t!]
	\begin{center}
		\includegraphics[width=0.49\textwidth]{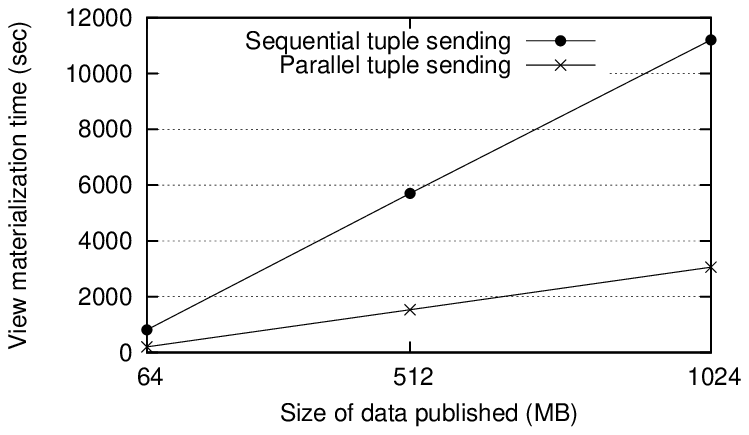}
	  	\includegraphics[width=0.49\textwidth]{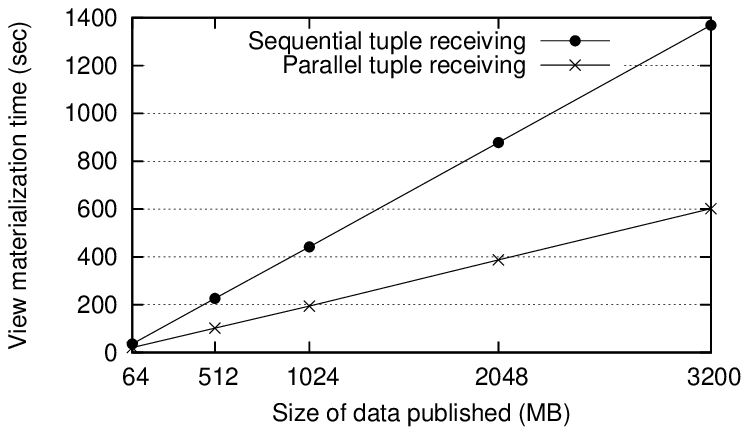}
	\vspace{-3mm}
		\caption{Experiment 4: one publisher, varying size of data, 64 consumers (left); experiment 5: 64 publishers, varying data size, one consumer (right).\label{fig:vary_data_many_consumers_many_producers}}
	\end{center}
\end{figure}

Like in Experiment 3, we run two variations of the same experiment: $(i)$ one for sequential tuple sending and $(ii)$ one for parallel tuple sending. The graph at left in Figure~\ref{fig:vary_data_many_consumers_many_producers} shows, as expected, that the materialization time increases linearly with the size of data published in the network in both cases. It also shows that the materialization time in the case of parallel tuple sending is considerably shorter (about 3000 sec. instead of 11500 sec. for absorbing 1024MBs of data).

\vspace{1mm}
\noindent\textbf{Experiment 5: 64 publishers, varying data size, one consumer}
We now study the potential for parallel publishing, i.e., the impact of the number of (simultaneous) publishers on the capacity of absorbing the data into a single view. The published data size varies from 64MBs to 3.2GBs, and  all the published data ends up in the view. The parameter values for this experiment are:

\vspace{-1mm}
\begin{center}
  \begin{tabular}[h!]{ c c c c c p{2mm} c p{2mm} c p{2mm} c p{2mm} }
    $|P|$ & $|P_{D}|$ & $|V|$ & $|P_{V}|$ &         $|D|$          & &  $|D_{V}|$              & & $|d|$ \\\hline
 	65   &    64     & 1     &    1      &   \{$64,\ldots,3200$\}   & &     \{$64,\ldots,3200$\}  & & 1MB \\
  \end{tabular}
\end{center}

Recall from Section~\ref{sec:archi} that the view materialization module maintains a queue of tuple-send requests and allows only a certain number of concurrent tuple-extractors to send data to it concurrently. In this experiment we test 2 modes of tuple-receiving concurrency: $(i)$ the consumer accepts only one tuple-send request at any given time (sequential tuple receiving); $(ii)$ the consumer accepts at most 64 tuple-send requests concurrently (parallel tuple receiving).

Figure~\ref{fig:vary_data_many_consumers_many_producers} (right) depicts the materialization time as the size of the published data increases.  We observe that the materialization time increases proportionally to the size of published data in both sequential and parallel tuple receiving modes. Also, parallel tuple receiving reduces the view materialization time by more than 50\% (600 sec. instead of about 1400 sec. to absorb 3.2GBs of data).

From the two graphs in Figure~\ref{fig:vary_data_many_consumers_many_producers}, we conclude that it is faster for the network to absorb data using one consumer and many publishers rather than many consumers and one publisher. For example, for absorbing 1024MBs of data, the view materialization time is less than 200 seconds (Figure~\ref{fig:vary_data_many_consumers_many_producers} right) for 64 publishers and one consumer, and about 3000 seconds in the case of one publisher and 64 consumers (Figure~\ref{fig:vary_data_many_consumers_many_producers} left). This is explained by the fact that data extraction is proven to be a slow process (Experiment 2) thus it is slow for a peer to extract all the available data by itself and ship them to the consumers.

\vspace{1mm}
\noindent\textbf{Experiment 6: varying number of publishers, fixed data, one consumer}
The purpose of this experiment is to study the impact that the parallelization of document publication has on the view materialization time. We use 250MBs of data distributed evenly across an increasing number of publishers. First, one peer publishes all the data, then two peers publish half of the data each, then 4, then 8 peers etc. The parameter values for this experiment are as follows:

\vspace{-1mm}
\begin{center}
  \begin{tabular}[h!]{ c p{2mm} c  p{2mm} c p{2mm} c p{2mm} c p{2mm} c p{2mm} c p{2mm} c p{2mm} }
    $|P|$ & & $|P_{D}|$ & & $|V|$ & & $|P_{V}|$ & &  $|D|$ & &  $|D_{V}|$ & & $|d|$ \\\hline
 	65  & &  \{1,2,\ldots,64\}  & &   1       & &    1   & &   512       & &     512     & & 0.49MB \\
  \end{tabular}
\end{center}

\begin{figure}[t!]
	\begin{center}
		\includegraphics[width=0.49\textwidth]{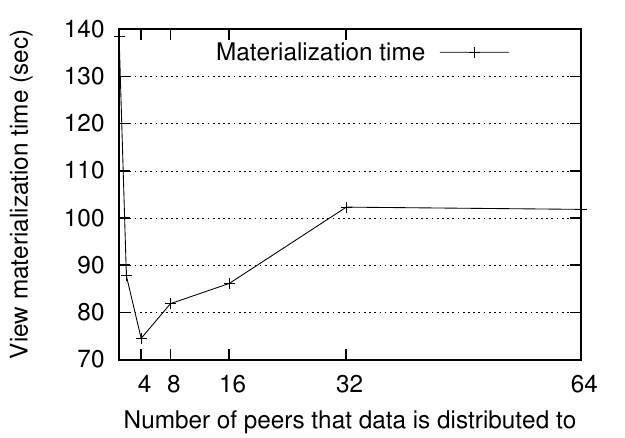}
		\includegraphics[width=0.49\textwidth]{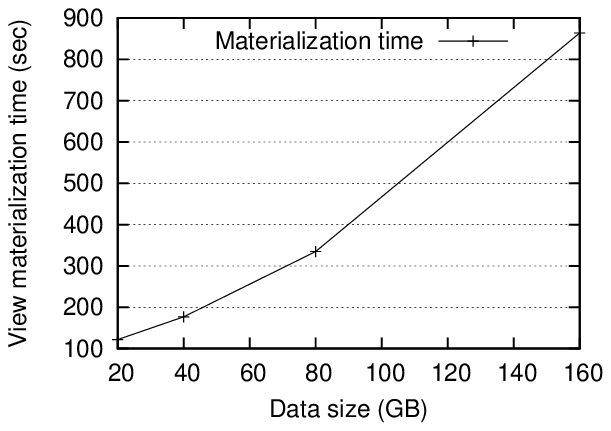}
\end{center}
\vspace{-5mm}
\caption{Experiment 6: publishing the same amount of data from an increasing number of publishers (left); experiment 7: publishing varying size of data in 50 groups of 5 peers each (right).\label{fig:one_view_many_publishers}}
\end{figure}

Figure~\ref{fig:one_view_many_publishers} (left) shows how materialization time varies depending on the number of parallel publishers. The time decreases as the data is distributed to two and then 4 publishers, as the extraction effort is parallelized. From 8 publishers onwards, the materialization time  increases again, until it stabilizes from 32 to 64 publishers. This increase is due to publishers simultaneously trying to connect to the consumer and making the consumer's storage module the bottleneck.

\vspace{1mm}
\noindent\textbf{Experiment 7: community publishing} \label{sec-simultaneous-materialization} We now consider a more complex scenario. We study materialization time in a setting with (logical) sub-networks, i.e., such that no single publisher has data of interest to all views, and no single view needs data from all publishers. The parameters of this experiment are:

\vspace{-4mm}
\begin{center}
  \begin{tabular}[h!]{ c c  c c c c c p{2mm} c p{2mm} }
    $|P|$ & $|P_{D}|$ & $|V|$ & $|P_{V}|$ & $|D|$ & & $|D_{V}|$ & & $|d|$ \\\hline
 	250   & 250 & 50 & 50 & \{$20K,\ldots,160K$\} & & \{$20K,\ldots,160K$\} & & 1MB \\
  \end{tabular}
\end{center}

We use a network of 250 peers, each of which holds the same number of 1MB documents. We logically divide the network into 50 groups of 5 peers each, such that in each group there are five publishers and one consumer (one peer is both a publisher and a consumer). The data of all publishers in a group is of interest for the consumer of that group, but it is not relevant for any of the other groups' views. The group peers are randomly chosen, i.e., they do not enjoy any special geographic or network locality etc. The total amount of data published (and shipped to the views) varies from 20GBs to 160GBs. Figure~\ref{fig:one_view_many_publishers} (right) shows that the materialization time grows linearly with the published data size.

\vspace{1mm}
\noindent\textbf{Conclusion} This Section has studied several extreme cases of view materialization (very skewed / very evenly distributed, with one or many publishers or consumers etc.), in order to traverse the space of possibilities. Overall, the experiments demonstrate the good scalability properties of ViP2P as the data volume increases, and that ViP2P exploits many parallelization opportunities when extracting, sending, receiving and storing view tuples. Table~\ref{tab:throughput} summarizes the results by providing a global metric, the view materialization throughput, reflecting the quantity of data that can be published (from documents to views) simultaneously in the network. Table~\ref{tab:throughput} demonstrates that ViP2P properly exploits all opportunities for parallelism in the ``community publishing'' scenario: the throughput is of 238 MB/s, while the best comparable result in this area from KadoP is of 0.33 MB/s only~\cite{DBLP:conf/icde/AbiteboulMPPS08}. 

\begin{table}[t!]
\begin{center}
  \begin{tabular}{ | p{0.045\columnwidth} | p{0.7\columnwidth} | p{0.14\columnwidth} | }
\hline
    Exp. No. & Experiment description  & Throughput (MB/sec) 	\\ \hline\hline
    3& One publisher, fixed data, varying number of consumers    &   \hfill	10.30	  		\\\hline 
    4& One publisher, varying data size, 64 consumers   &   \hfill	0.34	  		\\\hline 
	5& 64 publishers, varying data size, one consumer   &   \hfill	5.31	  		\\\hline 
	6& Varying number of publishers, fixed data, one consumer   &  \hfill 	8.05	  		\\\hline 
	7& Community publishing  &  \hfill 	238.80	  		\\\hline 
\hline
  \end{tabular}
\vspace{-2mm}
\caption{Maximum data absorption throughput during view materialization.\label{tab:throughput}}
\end{center}
\vspace{-2mm}
\end{table}

\Subsection{View indexing and retrieval evaluation}
\label{sec:experimental-view-indexing-retrieval}

We now compare the view indexing and lookup strategies LI, RLI, LPI and RPI described in Section~\ref{sec:index}.

\vspace{2mm}
\noindent\textbf{Experiment 8: view indexing and retrieval} We start with a random synthetic query $q$ of height 5, having  $30$ nodes labeled $a_1,\ldots, a_{30}$. Each node of $q$ has between 0 and 2 children.
We then create three variants of $q$:

\vspace{-2mm}
\begin{itemize}
\item $q'$  has the same labels as $q$, but totally disagrees with $q$ on the structure (if $a_i$ is an ancestor of $a_j$ in $q$, $a_i$ is not an ancestor of $a_j$ in $q'$)
\item  $q''$ coincides with $q$ for half of the query, while the other half conserves the labels of $q$ but totally disagrees on the structure (as in $q'$)
\item $q'''$  has the same structure as $q$, half of it has the same labels $a_1,\ldots,a_{15}$, while the other half uses a different set of labels $b_1,\ldots,b_{15}$ (that replace $a_{16},\ldots,a_{30}$ respectively).
\end{itemize}
\vspace{-2mm}

From each of $q$, $q'$, $q''$ and $q'''$ we automatically generate $360$ views of 2 to 5 nodes, totaling $1440$ views, such that: the views can all be embedded into their respective queries, i.e. those generated from $q$ can be embedded in $q$, those generated from $q'$ can be embedded in $q'$ and so on. We, thus, obtain a mix of views resembling the original query $q$ to various degrees. 

We have indexed the resulting $1440$ views in a network of 250 peers, following the LI, RLI, LPI and RPI strategies described in Section~\ref{sec:index}. We then performed lookups using the four different indexing strategies. The parameters characterizing this experiment are the following:

\vspace{-1mm}
\begin{center}
  \begin{tabular}[h!]{ c p{2mm} c  p{2mm} c p{2mm} c p{2mm} c p{2mm} c p{2mm} c p{2mm} c p{2mm} }
    $|P|$ & & $|P_{D}|$ & & $|V|$ & & $|P_{V}|$ & & $|D|$ & & $|D_{V}|$ & & $|d|$ \\\hline
 	250 & & 0 & & 1440 & & 250 & & 0 & & 0 & & 0 \\
  \end{tabular}
\end{center}

\begin{figure}[t!]
  \begin{center}
	\scalebox{1}{
    	\includegraphics[width=0.49\textwidth]{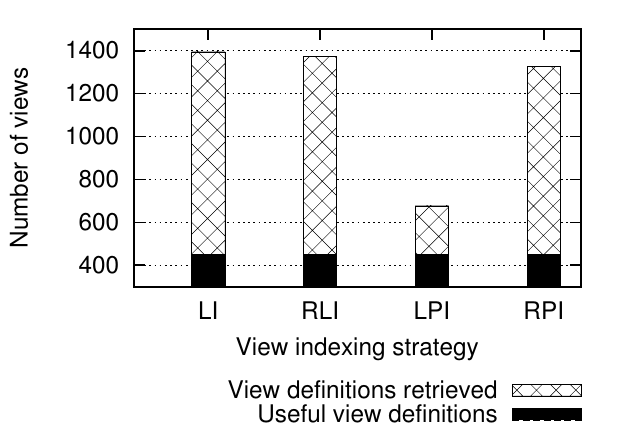}
	    \includegraphics[width=0.49\textwidth]{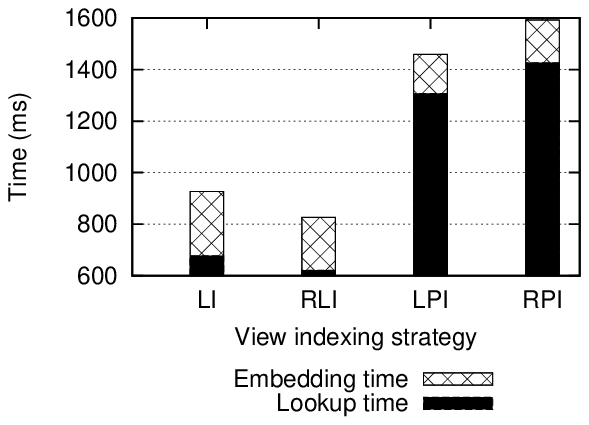}
  	}
\vspace{-6mm}
  \caption{Experiment 8: view definition retrieval (left); embedding vs lookup time (right).} \label{fig:view_definition_lookups}
\end{center}
\vspace{-6mm}
 \end{figure}

Figure~\ref{fig:view_definition_lookups} (left) depicts the number of views retrieved by each strategy, compared to the number of useful views, which can be embedded into $q$.  We observe, as expected, that the path indexing-lookup strategies (LPI and RPI) are more precise than the label based ones (LI and RLI). Moreover, LPI is the most precise, since it uses as keys longer paths, describing views more precisely.

\begin{figure}[t!]
  \begin{center}
	\scalebox{1}{
    	\includegraphics[width=0.49\textwidth]{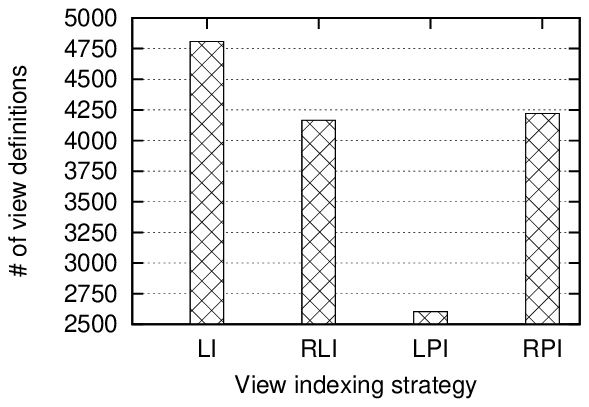}
	    \includegraphics[width=0.49\textwidth]{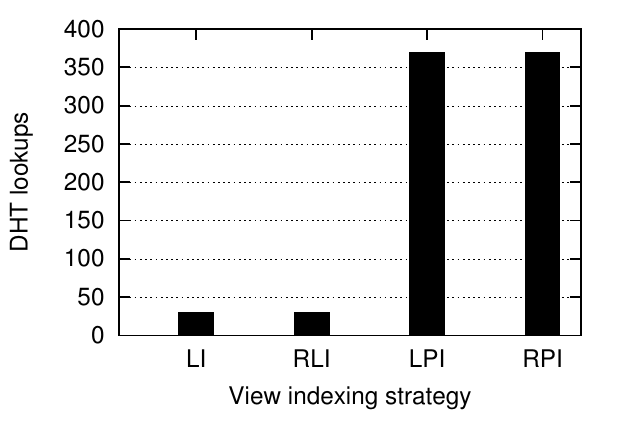}
  	}
\vspace{-8mm}
  \caption{Experiment 8: lookups generated for retrieving views (left); embedding vs lookup time (right).} \label{fig:view_definition_lookups2}
\end{center}
\vspace{-2mm}
 \end{figure}

Figure~\ref{fig:view_definition_lookups} (right) depicts the time spent looking up in the DHT the set of (possibly) useful views in order to rewrite $q$, as well as the time spent to check whether embeddings exist from those views into $q$. We observe that from this angle, the label strategies (LI and RLI) perform better than the path strategies, since the more numerous lookups performed by the path strategies take up too much time when processing queries.

Figure~\ref{fig:view_definition_lookups2} (left) depicts the number of view definitions that were indexed in the DHT by each view indexing strategy. Figure~\ref{fig:view_definition_lookups2} (right) depicts the number of lookups performed by each strategy for the query we consider. As expected, LI inserts the largest number of DHT entries. With respect to query-driven lookup, LI and RLI perform 30 lookups, much less than LPI and RPI that perform 370 lookups each. 

From this experiment, we conclude that label-based strategies are preferable, since the savings at query processing time are more critical than the DHT index size (which is very modest in all cases) or the precision of look-up, as the retrieved view definitions are further filtered at the query peer (after the embedding filtering, the rewriting is run with the same set of views no matter the used strategy).

\Subsection{Query engine evaluation}
\label{sec-exp-query}

\noindent\textbf{Experiment 9: query response time vs. query selectivity and number of results} 
We now investigate the query processing performance as the data size increases. We use 20 peers, all of which are publishers, 2 are consumers and 1 is a query peer.  The query peer and the 2 consumers are located in 3 different locations of France (Bordeaux, Lille and Orsay). The parameter values characterizing this experiment are the following:

\vspace{-1mm}
\begin{center}
  \begin{tabular}[h!]{ c c  c c c p{2mm} c p{2mm} c p{2mm} c p{2mm} }
    $|P|$ & $|P_{D}|$ & $|V|$ & $|P_{V}|$ & $|D|$             & & $|D_{V}|$         & & $|d|$ \\\hline
 	20    &   20     &   2   &    2      & \{$20,\ldots,500$\} & & \{$20,\ldots,500$\} & & 0.5MB \\
  \end{tabular}
\end{center}

The document used in this experiment is the same as the one of Figure~\ref{fig:64-tag-document_and_views_per_peer} (left) with a slight difference: the root element $catalog$ has only one child, named $camera$. \\

\noindent The views defined in the network are the following:
\vspace{-1mm}
\begin{itemize}
	\item $v_1$ is \kwd{//$catalog_{ID}$//$camera_{ID}$//$description_{ID,cont}$}
	\item $v_2$ is \kwd{//$catalog_{ID}$//$camera_{ID}$//\{$description_{ID}$,~$price_{ID,val}$,~$specs_{ID,cont}$\}}
\end{itemize}

\noindent Each view stores one tuple from each document. A $v_1$ tuple from document $d$ roughly contains all of $d$ (since the $description$ element is the most voluminous in each $camera$). A $v_2$ tuple is quite smaller since it does not store the full camera descriptions. We use three queries:

\vspace{-1mm}
\begin{itemize}
	\item $q_1$ asks for the $description_{cont}$, $specs_{cont}$ and $price_{val}$ of each $camera$. To evaluate $q_1$, ViP2P joins the views $v_1$ and $v_2$. Observe that $q_1$ returns full XML elements, and in particular, product descriptions, which are voluminous in our data set. Therefore, $q_1$ returns roughly all the published data (from 10MB in 20 tuples, to 250MB in 500 tuples). 
	\item $q_2$ requires the $description_{ID}$, $specs_{ID}$ and $price_{ID}$ of each $camera$. This is very similar to $q_1$ but it can be answered based on $v_2$ only. The returned data is much smaller since there are only IDs and no XML elements: from 2KB in 20 tuples, to 40KB in 500 tuples.
     \item $q_3$ returns the $specs//sensor\_type_{val}$ of each $camera$. The rewriting of $q_3$ applies \emph{navigation} over $specs_{cont}$ that is stored by $v_2$. The result size varies from 2KB in 20 tuples to 40KB in 500 tuples.
\end{itemize}

\begin{figure}[t!]
  \begin{center}
   	\scalebox{1.02}{
 		\includegraphics[width=0.32\textwidth]{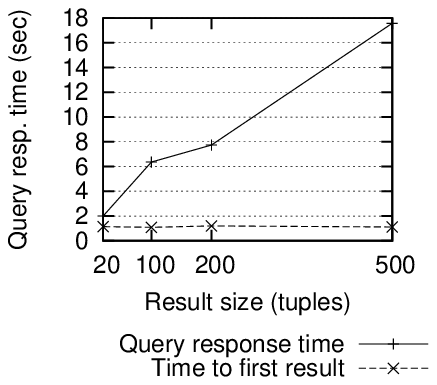}
	    \includegraphics[width=0.32\textwidth]{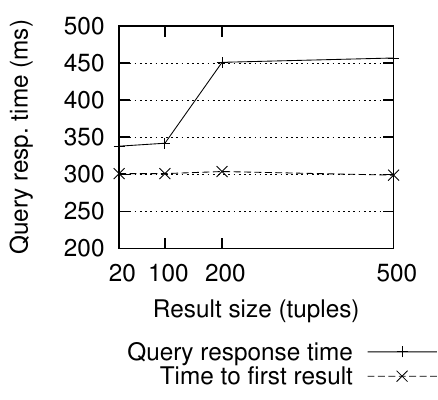}
	    \includegraphics[width=0.32\textwidth]{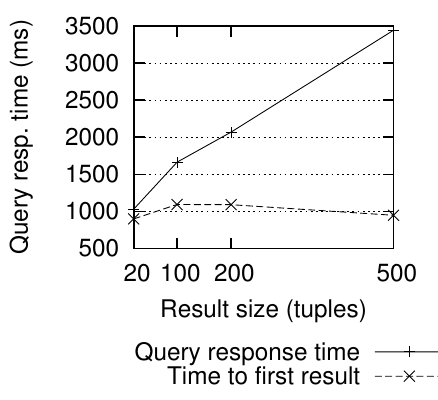}
	}
  \end{center}
	\vspace{-6mm}
  \caption{Experiment 9: query execution time vs. number of result tuples for three queries.\label{query_execution_time_vs_resutls}}
\end{figure}

Figure~\ref{query_execution_time_vs_resutls} shows the query response time and the time to get the first result for the 3 queries. 
The low selectivity query $q_1$ (at left in Figure~\ref{query_execution_time_vs_resutls}) takes longer than $q_2$, due to the larger data transfers and the necessary view join. The time to first result is always constant for both $q_1$ and $q_2$ and does not depend on the result size. For $q_1$, a hash join is used to combine $v_1$ and $v_2$, and thus no tuple is output before the view $v_2$ has been built into the buckets of the hash join. This is done in more or less one second in the case of $q_1$ and about 300ms for $q_2$. Note that the join is performed on the peer holding $v_1$ as it is faster to transfer $v_2$ at the peer holding $v_1$. Increases in the total running time appear when more data-sending messages are needed to transfer increasing amounts of results. For $q_3$, which applies navigation on the view $v_2$, the time to the first tuple is the time to evaluate the navigation query locally at $v_2$'s peer and send the first message with result tuples to the query peer, and this does not grow with the data size.

\vspace{1mm}
\noindent\textbf{Conclusion} The ViP2P query processing engine scales close to linearly when answering queries in a wide-area network. The fact that ViP2P rewrites queries into logical plans which are then passed to an optimizer enables it to take advantage of known optimization techniques used in XML and/or distributed databases, to reduce the total query evaluation time, and (depending on the characteristics of the particular physical operators chosen) the time to the first answer. Given the ViP2P architecture, the peers involved in processing a query are only those holding the views used in the query rewriting; this is why using only 20 peers for this experiment does not affect its interpretation, since ViP2P query processing involves only three peers. The network size may only impact the view look-up time, which is very modest (Section~\ref{sec:experimental-view-indexing-retrieval}).

\Subsection{Conclusion of the experiments}

Our study leads to the conclusion that the ViP2P architecture scales up well. In particular, view materialization scales in the number of publishers and consumers, in the size of the network, and in the size of the data.
High contention at a single consumer receiving data from many publishers, and especially at a single publisher contributing to many consumers' views, degrades the ability of the view holders to efficiently absorb data. However, these contention effects are to be expected in a large distributed system. 
Moreover, we showed that when interest in the published data is more evenly distributed among sub-communities, ViP2P takes advantage of all parallelization opportunities to increase the data transfer rate between publishers and consumers by 3 orders of magnitude. 
Our view materialization experiments also show the importance of carefully tuning all stages in the data extraction and data transfer process, including asynchronous communication and parallelization whenever possible. The cumulated impact of these optimizations on the data transfer rate between peers are dramatic (more than 4 orders of magnitude increase).

Our query processing experiments  show that label-based view indexing strategies are preferable, and indeed we use RLI by default. They also demonstrate that the ViP2P execution engine scales linearly up to large data volumes, orders of magnitude more than in previous real DHT deployments~\cite{DBLP:conf/icde/AbiteboulMPPS08,10.1109/TKDE.2009.26}.

%% file: conclusion.tex
\Section{Conclusion and perspectives}
The efficient management of large XML corpora in structured P2P networks requires the ability to deploy data access support structures, which can be tuned to closely fit application needs. We have presented the VIP2P approach for building and maintaining structured materialized views, and processing peer queries based on the existing views in the DHT network. Using DHT-indexed views adds to query processing the (modest) cost of locating relevant views and rewriting the query using the views, in exchange for the benefits of using pre-computed results stored in views.  We studied several view indexing strategies and associated complete view lookup methods. Moreover, we did an extensive study of our platform's main aspects (view materialization, indexing and retrieval, and query processing) in different scenarios and settings. ViP2P  was able to extract and disseminate 160GB of data in less than 15 minutes over 250 computers in a WAN network~\cite{Grid5000}. These results largely improve over the closest competing XML management platforms based on DHTs, and actually implemented and deployed (1 GB of data indexed in 50 minutes in KadoP~\cite{DBLP:conf/icde/AbiteboulMPPS08}, hundreds of MB of data on 11 peers in psiX~\cite{10.1109/TKDE.2009.26}, which, unlike us, focused only on document indexing and look-up). 

Many avenues for further research are open. An ongoing work built on ViP2P, LiquidXML~\cite{LiquidXML} automatically selects and continuously adapts a set of materialized views on each peer, to improve query processing performance in the network. Handling documents that contain references to each other and evaluating tree pattern queries that extend to many documents are other interesting developments.

\vspace{2mm}
\noindent {\bf Acknowledgements} Part of the ViP2P code comes from ULoad~\cite{DBLP:conf/vldb/ArionBMP07}. We thank Alin Tilea, Jes\'us Camacho-Rodr\'iguez, Alexandra Roatis, Varunesh Mishra and Julien Leblay for their help developing and testing ViP2P. Experiments presented in this paper were carried out using the Grid'5000 experimental testbed, being developed under the INRIA ALADDIN development action with support from CNRS, RENATER and several Universities as well as other funding bodies (see https://www.grid5000.fr). 